\documentclass[12pt,a4paper]{article}

\usepackage[a4paper,margin=1in]{geometry}
\usepackage{amsmath,amssymb,amsthm,mathtools}
\usepackage{enumitem}
\usepackage{array,booktabs,longtable}
\usepackage{tikz}
\usepackage{xcolor}
\usepackage{hyperref}
\usepackage{setspace}
\usepackage{microtype}
\usepackage[numbers,sort&compress]{natbib}

\hypersetup{colorlinks=true,linkcolor=blue,citecolor=blue,urlcolor=blue}
\newtheorem{theorem}{Theorem}[section]
\newtheorem{proposition}[theorem]{Proposition}

\theoremstyle{definition}
\newtheorem{definition}[theorem]{Definition}

\newcommand{\OO}{\mathcal O}
\newcommand{\Rnn}{\mathbb R_{\ge 0}}

\title{Operational Regimes of Ternary $\Gamma$-Semiring\\
Chemical Systems}

\author{\begin{tabular}{c}
Chandrasekhar Gokavarapu$^{1,2,*}$ \\
Venkata Rao Kaviti$^{3}$ \\
Srinivasa Rao Tirunagari$^{3}$ \\
Dr Sajani Lavanya Madasi$^{1}$ \\
Dr Madhusudhana Rao Dasari$^{4,5}$
\end{tabular}}
\date{}

\begin{document}
\maketitle

\begin{center}
\small
$^{1}$Department of Mathematics, Government College (Autonomous), Rajahmundry, Andhra Pradesh, India, PIN 533105\\
$^{2}$Department of Mathematics, Acharya Nagarjuna University, Guntur, Andhra Pradesh, India, PIN 522510\\
$^{3}$Department of Chemistry, Government College (Autonomous), Rajahmundry, Andhra Pradesh, India, PIN 533105\\
$^{4}$Department of Mathematics, Government College for Women (Autonomous), Guntur, Andhra Pradesh, India, PIN 522001\\
$^{5}$Department of Mathematics, Acharya Nagarjuna University, Guntur, Andhra Pradesh, India, PIN 522510\\[2pt]
$^{*}$Corresponding author: chandrasekhargokavarapu@gmail.com; chandrasekhargokavarapu@gcrjy.ac.in
\end{center}

\begin{abstract}
Ternary $\Gamma$-semiring chemical systems provide an algebraic language in which mediators such as catalysts, solvents, inhibitors, thermodynamic conditions, and external fields enter the transformation law as internal arguments.  Building on the published MATCH foundation article, this paper studies operational regimes of the compressed TGS notation rather than proposing a second axiomatization.  Distributivity is interpreted as observable non-interference of mediated channels, while associativity is interpreted as observable path-independence of bracketed mediated protocols.  Since the state space is only semigroup- or semiring-level, subtraction is not used at the state level; instead, observable defect and ratio indices are defined after applying a non-negative observable.  We prove that distributivity implies observable non-interference under observable additivity, and that associativity implies observable path-independence.  Nonzero defects detect observable distinguishability.  To strengthen the diagnostic layer, we prove recovery theorems showing that separating families of observables can recover algebraic distributivity and associativity from vanishing observable defects.  We also introduce approximate operational regimes and prove ratio bounds controlled by uniform defect estimates.  Under strict convexity, uniqueness, and common global-constraint assumptions, a potential-generated equilibrium operation is shown to be associative.  A symbolic intermediate-dependent Michaelis--Menten type construction shows how non-associativity may arise in kinetic regimes, and regulatory/allosteric-type non-distributivity is discussed cautiously through observable coupling.
\end{abstract}



\medskip
\begin{footnotesize}
\noindent\textbf{Author contact and profile information.}
Chandrasekhar Gokavarapu: \href{mailto:chandrasekhargokavarapu@gmail.com}{chandrasekhargokavarapu@gmail.com}; \href{mailto:chandrasekhargokavarapu@gcrjy.ac.in}{chandrasekhargokavarapu@gcrjy.ac.in}; \href{https://gcrjy.ac.in/facultyprofile.php?facultyid=204&departmentid=44}{Institutional profile}; \href{https://scholar.google.com/citations?user=8EQ6oz8AAAAJ&hl=en}{Google Scholar}; \href{https://www.researchgate.net/profile/Chandrasekhar-Gokavarapu}{ResearchGate}; \href{https://orcid.org/0009-0006-5306-371X}{ORCID: 0009-0006-5306-371X}.
Venkata Rao Kaviti: \href{mailto:venkatkaviti@gmail.com}{venkatkaviti@gmail.com}; \href{https://orcid.org/0000-0001-7229-2899}{ORCID: 0000-0001-7229-2899}.
Srinivas Thirunagari: \href{mailto:tsssrinu@gcrjy.ac.in}{tsssrinu@gcrjy.ac.in}; \href{https://orcid.org/0009-0003-1615-2199}{ORCID: 0009-0003-1615-2199}.
Sajani Lavanya Madasi: \href{mailto:sajanimaths@gcrjy.ac.in}{sajanimaths@gcrjy.ac.in}; \href{https://orcid.org/0009-0000-8194-4027}{ORCID: 0009-0000-8194-4027}.
Madhusudhana Rao Dasari: \href{mailto:dmrmaths@gmail.com}{dmrmaths@gmail.com}; \href{https://orcid.org/0000-0003-0897-868X}{ORCID: 0000-0003-0897-868X}.
\end{footnotesize}

\section{Introduction}

Chemical transformations are rarely determined only by the formal identities of the species appearing in a reaction scheme.  In many situations, the observed outcome also depends on mediating data such as catalysts, solvents, inhibitors, temperature--pressure regimes, pH conditions, or external fields.  These mediators may change accessible pathways, suppress competing routes, stabilize intermediates, or select between equilibrium and non-equilibrium outcomes.  This makes mediated chemical transformation a natural setting for algebraic models in which the reaction law is not merely binary but depends intrinsically on additional operational parameters.

The axiomatic foundation for such a viewpoint was introduced in the published MATCH article \cite{GokavarapuMATCH2026Axiomatic}, where chemical systems were modelled as ternary $\Gamma$-semirings and the symbols in the state set and parameter set were interpreted as chemical states and mediating conditions, respectively.  That article developed the basic language of TGS-chemical systems, including ideals, $\Gamma$-ideals, prime and semiprime chemical ideals, reaction pathways, and homomorphisms.  The present paper is not a second axiomatization of TGS-chemical systems. Rather, it studies the operational consequences and failure regimes of the axioms introduced previously.

The remaining gap is operational rather than purely axiomatic.  Although ternary $\Gamma$-semiring models provide a formal language for mediated chemical transformations, the observable meaning of associativity and distributivity has not yet been systematically separated from regimes where these identities fail.  In particular, equilibrium-like situations may display path-independent behaviour, while kinetic, inhibited, cooperative, regulatory, or allosteric regimes may exhibit mediator-sensitive deviations from strict associativity or distributivity.  Classical kinetic and thermodynamic theories already show that catalytic and environmental conditions enter reaction behaviour in a structured way \cite{MichaelisMenten1913Invertinwirkung,BriggsHaldane1925EnzymeKinetics,Arrhenius1889ReactionVelocity,Eyring1935ActivatedComplex,Alberty2003ThermodynamicsBiochemicalReactions}; the purpose here is to translate such operational distinctions into the TGS setting without claiming that every chemical regime must satisfy the exact algebraic axioms.

The contributions of this paper are as follows.  First, we give a notation bridge from expanded chemical notation, where states and mediators are written explicitly, to compressed TGS notation.  Second, we interpret distributivity as an observable non-interference condition between mediated channels.  Third, we interpret associativity as observable path-independence of iterated mediated transformations.  Fourth, we introduce observable defect and ratio indices measuring deviations from exact associativity or distributivity.  Fifth, we prove recovery theorems showing that separating families of observables can recover algebraic distributivity and associativity from vanishing observable defects.  Sixth, we introduce approximate operational regimes and prove ratio estimates controlled by uniform defect bounds.  Seventh, we prove a potential-generated equilibrium theorem giving a sufficient condition under which path-independence follows from an underlying scalar potential.  Eighth, we construct a symbolic intermediate-dependent Michaelis--Menten type example showing how non-associativity may arise in sequential kinetic regimes \cite{JohnsonGoody2011OriginalMichaelisConstant,SchnellMendoza1997ClosedForm,CornishBowden2015OneHundredYears,Bisswanger2017EnzymeKinetics}.  Ninth, we discuss regulatory and allosteric-type non-distributivity, motivated by standard cooperative binding models \cite{Hill1910Haemoglobin,MonodWymanChangeux1965AllostericTransitions,KoshlandNemethyFilmer1966BindingModels}.  Finally, we summarize the operational outcomes in a fourfold regime classification table distinguishing exact TGS regimes, approximately TGS regimes, defect-measurable regimes, and genuinely non-TGS regulatory regimes.

The paper is organized as follows.  Section~2 fixes notation and recalls only the minimum TGS background needed from the published MATCH foundation article \cite{GokavarapuMATCH2026Axiomatic}, together with standard algebraic sources on semirings, ternary operations, and parameterized algebraic structures \cite{Golan1999SemiringsApplications,HebischWeinert1998Semirings,Post1940PolyadicGroups,Lister1971TernaryRings,Barnes1966GammaRingsNobusawa,Tropashko2006TernaryRelationsSemiring}.  Section~3 introduces the core operational laws, observables, and defect and ratio indices.  Sections~4 and~5 develop distributivity as non-interference and associativity as path-independence.  Section~6 proves algebraic recovery theorems for separating observable families.  Section~7 develops approximate operational regimes and ratio estimates under small defect bounds.  Section~8 proves the potential-generated equilibrium theorem.  Section~9 gives a symbolic intermediate-dependent Michaelis--Menten type non-associative construction, using enzyme-kinetic references only as conceptual motivation rather than as experimental data \cite{MichaelisMenten1913Invertinwirkung,BriggsHaldane1925EnzymeKinetics,JohnsonGoody2011OriginalMichaelisConstant,SchnellMendoza1997ClosedForm}.  Section~10 discusses regulatory and allosteric-type non-distributivity \cite{Hill1910Haemoglobin,MonodWymanChangeux1965AllostericTransitions,KoshlandNemethyFilmer1966BindingModels}.  Section~11 records the fourfold operational classification table, Section~12 states limitations and scope, and Section~13 concludes the paper.
\section{Background and Notation Bridge}

This section fixes the notation used throughout the paper.  The purpose is not to repeat the full axiomatic development of TGS-chemical systems from \cite{GokavarapuMATCH2026Axiomatic}, but to introduce the compressed operational notation needed for the present study.  The notation below is deliberately economical: it records states, mediators, and mediated outputs, while leaving room for measuring the failure or validity of associativity and distributivity in later sections.

\subsection{Compressed TGS-chemical systems}

\begin{definition}[Compressed TGS-chemical system]
A compressed TGS-chemical system is a quadruple
\[
(S,\Gamma,+,\circ),
\]
where $(S,+)$ is a commutative semigroup of chemical states, $\Gamma$ is a nonempty mediator set, and
\[
\circ:S\times \Gamma\times S\longrightarrow S,
\qquad (a,\gamma,b)\mapsto a\gamma b,
\]
assigns to two states $a,b\in S$ and a mediator $\gamma\in\Gamma$ a mediated output state $a\gamma b\in S$.
\end{definition}

The operation $\circ$ will usually be written without the symbol $\circ$; thus
\[
\circ(a,\gamma,b)=a\gamma b.
\]
This notation is standard in the theory of $\Gamma$-type algebraic structures, where the parameter is treated as an internal argument of the operation rather than as an external label.  It is also compatible with the ternary-operation viewpoint used in polyadic and ternary algebra \cite{Post1940PolyadicGroups,Lister1971TernaryRings,Barnes1966GammaRingsNobusawa}.  The semigroup operation $+$ records formal aggregation or parallel combination of coarse-grained states, while the mediated operation $a\gamma b$ records a directed operational transformation under the mediator $\gamma$.

\subsection{Chemical interpretation of the objects}

The set $S$ is interpreted as a collection of coarse-grained chemical states.  Depending on the level of modelling, an element of $S$ may represent a molecular species, a reaction intermediate, a phase descriptor, a concentration class, an electronic state, or an experimentally distinguishable macro-state.  The use of a coarse-grained state set is intentional: the present article is not a molecular simulation model, nor does it attempt to encode all microscopic degrees of freedom.  Instead, it studies structural laws that may hold, approximately hold, or fail for mediated transformations between such states.

The mediator set $\Gamma$ represents contextual data that influence the transformation rule.  Typical elements of $\Gamma$ may encode catalysts, solvents, thermodynamic conditions, inhibitors, pH regimes, pressure--temperature environments, ionic strength, external electromagnetic fields, or regulatory constraints.  In kinetic examples, a mediator may summarize rate-changing information; in thermodynamic examples, it may summarize equilibrium conditions; and in regulatory examples, it may summarize inhibition or cooperative control.  Classical chemical kinetics and thermodynamics already show that such contextual data can determine observable outcomes \cite{MichaelisMenten1913Invertinwirkung,BriggsHaldane1925EnzymeKinetics,Arrhenius1889ReactionVelocity,Eyring1935ActivatedComplex,Alberty2003ThermodynamicsBiochemicalReactions}.  The present framework treats that contextual dependence algebraically.

Thus, $a\gamma b$ should not be read as an ordinary binary reaction between $a$ and $b$ with $\gamma$ written above an arrow as a passive annotation.  Rather, $\gamma$ is part of the input data of the operation itself.  Two expressions
\[
a\gamma_1 b
\qquad\text{and}\qquad
a\gamma_2 b
\]
may represent different operational outcomes even when the two state entries $a$ and $b$ are unchanged.  This distinction is essential for modelling catalyst-dependent, solvent-dependent, inhibitor-dependent, field-dependent, and environment-dependent transformations in a single structural language.

\subsection{Bridge from expanded notation to compressed notation}

The published MATCH foundation article used the expanded notation
\[
[A,\alpha,B,\beta,C]
\]
to represent a mediated ternary transformation involving three chemical states $A,B,C\in S$ and two mediators $\alpha,\beta\in\Gamma$ \cite{GokavarapuMATCH2026Axiomatic}.  In the present paper we use the compressed convention
\[
[A,\alpha,B,\beta,C]=(A\alpha B)\beta C.
\]
This convention interprets the expanded expression as an iterated mediated operation: first the state $A$ interacts with the state $B$ under the mediator $\alpha$, producing the intermediate compressed output $A\alpha B$; then this output interacts with $C$ under the mediator $\beta$.  The compressed notation is used only for operational diagnostics in this paper; it is not intended to replace the more general expanded TGS formalism of \cite{GokavarapuMATCH2026Axiomatic}.

The expression $a\gamma b$ is ternary in the algebraic sense because it is the value of a map
\[
S\times \Gamma\times S\longrightarrow S.
\]
It has two state arguments and one mediator argument.  Therefore the notation is not merely binary reaction notation.  The mediator is an internal argument of the operation, and changing the mediator changes the operation being applied.  This point is important because the present paper studies exactly those cases in which changing mediators, changing the order of mediated composition, or separating parallel mediated channels changes the observed result.

Under this convention, an expression such as
\[
(a\alpha b)\beta c
\]
represents one parenthesized operational route, while
\[
a\alpha(b\beta c)
\]
represents another.  The equality
\[
(a\alpha b)\beta c=a\alpha(b\beta c)
\]
is therefore not a notational tautology.  It is an observable path-independence condition.  Similarly, distributive identities involving $+$ are not formal decorations; they express conditions under which mediated transformation does not interfere with formal aggregation or parallel combination of states.  These two interpretations are developed in Sections~4 and~5.

\subsection{No-duplication boundary with the accepted article}

Only the minimum background needed for the operational theory is recalled here.  The foundational paper \cite{GokavarapuMATCH2026Axiomatic} introduced the general TGS-chemical viewpoint, interpreted state and mediator sets, developed chemical ideals and $\Gamma$-ideals, studied prime and semiprime chemical ideals, and discussed homomorphisms and reaction pathways.  Those constructions are not repeated in this article except where they are needed to motivate notation.

The present paper changes the emphasis from foundational axiomatization to operational diagnosis.  Its main questions are the following:
\[
\text{When does a mediated expression have path-independent value?}
\]
\[
\text{When does a mediated operation distribute over parallel state-combination?}
\]
\[
\text{When these laws fail, can the failure be measured structurally?}
\]
Accordingly, we use standard algebraic terminology from semiring and ternary-operation theory only as background \cite{Golan1999SemiringsApplications,HebischWeinert1998Semirings,Post1940PolyadicGroups,Lister1971TernaryRings,Tropashko2006TernaryRelationsSemiring}.  The new content begins from the operational interpretation of associativity, distributivity, and their observable defect indices, not from a second definition of TGS-chemical systems.

This boundary is important for the logic of the article.  The MATCH foundation article supplies the structural setting.  The present paper asks how that setting behaves in equilibrium-like, kinetic, inhibited, cooperative, and regulatory regimes.  Exact associativity and exact distributivity will be treated as special operational regimes, while deviations from them will be treated as meaningful failure modes rather than as mere violations of formal axioms.
\section{Operational Laws and Observable Diagnostics}

This section records the operational laws studied in the rest of the paper and introduces observable diagnostics for regimes in which the exact algebraic laws may fail.  The point is not to assume that every mediated chemical system is exactly associative or exactly distributive.  Rather, associativity and distributivity are treated as testable structural regimes, while their failures are measured through observables.  This is important for applications to kinetic, inhibited, cooperative, and regulatory settings, where exact algebraic identities may be too strong but their deviations may still carry meaningful chemical information \cite{MichaelisMenten1913Invertinwirkung,BriggsHaldane1925EnzymeKinetics,MonodWymanChangeux1965AllostericTransitions,KoshlandNemethyFilmer1966BindingModels}.

\subsection{Core algebraic laws}

\begin{definition}[Associativity]
A TGS-chemical system is associative on the indicated states and mediators if
\[
(a\gamma b)\mu c=a\gamma(b\mu c).
\]
\end{definition}

Associativity says that two mediated routes with the same ordered input data produce the same final state.  The left-hand side first combines $a$ and $b$ under $\gamma$, and then combines the result with $c$ under $\mu$.  The right-hand side first combines $b$ and $c$ under $\mu$, and then combines $a$ with that result under $\gamma$.  Thus, in the present operational interpretation, associativity is a path-independence condition.  It is not merely a formal algebraic identity; it asserts that two different parenthesized mediated pathways are indistinguishable at the state level.

\begin{definition}[Right and left distributivity]
Right distributivity means
\[
a\gamma(b+c)=a\gamma b+a\gamma c,
\]
and left distributivity means
\[
(a+b)\gamma c=a\gamma c+b\gamma c.
\]
\end{definition}

Right distributivity says that applying the mediated transformation to a parallel or aggregated input $b+c$ gives the same result as transforming $b$ and $c$ separately and then aggregating the outputs.  Left distributivity has the analogous meaning in the first state position.  Operationally, distributivity is interpreted as a non-interference law: the channels represented by $b$ and $c$ do not inhibit, enhance, or otherwise modify one another under the mediator $\gamma$.  This interpretation is consistent with the role of distributive laws in semiring theory, where addition and multiplication encode parallel and sequential composition \cite{Golan1999SemiringsApplications,HebischWeinert1998Semirings,Tropashko2006TernaryRelationsSemiring}.

\subsection{Subtraction safety}

Since $S$ is only assumed to be a semigroup or semiring-level object, subtraction is not a primary operation on $S$.  Therefore, expressions such as
\[
(a\gamma(b+c))-(a\gamma b+a\gamma c)
\]
are not meaningful inside $S$ unless additional structure is explicitly imposed.  In particular, algebraic defects should not be defined by subtraction at the state level unless $S$ has been embedded into a group completion, an ordered vector space, a normed space, or another structure equipped with a valid difference operation.

This restriction is important for mathematical safety.  A semiring or TGS-semiring framework is designed precisely to allow non-negative, irreversible, or aggregation-based systems where additive inverses may not exist.  Chemical quantities such as concentrations, extents, yields, intensities, and heat magnitudes are commonly non-negative.  Hence it is more natural to measure deviations after applying an observable map into a numerical space rather than by subtracting states directly.

Accordingly, this paper defines defect indices only after applying observables.  The observable level is where subtraction, absolute value, ratios, and numerical comparison become legitimate.

\subsection{Observable maps}

\begin{definition}[Observable]
An observable is a map
\[
\OO:S\longrightarrow \Rnn,
\]
chosen according to the modelling context.
\end{definition}

The observable $\OO$ converts a coarse-grained state into a non-negative numerical quantity.  It may represent a directly measured experimental value, a model-derived scalar, or a structural score assigned to a state.  No universal observable is assumed.  The correct choice of $\OO$ depends on the chemical question being asked and on the level of coarse-graining used in the state space $S$.

An observable may fail to distinguish different states.  Thus, $\OO(x)=\OO(y)$ does not necessarily imply $x=y$.  This distinction will be important below: observable associativity is weaker than algebraic associativity unless $\OO$ separates the relevant states.

\subsection{Examples of observables}

Typical observables include product concentration, reaction extent, heat release, absorbance, fluorescence intensity, conversion ratio, catalytic turnover, equilibrium yield, or a normalized activity measure.  For example, in an enzyme-kinetic setting, $\OO(x)$ may record product concentration or reaction velocity associated with the state $x$, while in a thermodynamic setting it may record heat release or a potential-derived scalar.  In optical or spectroscopic settings, $\OO(x)$ may represent absorbance or fluorescence intensity.

Such choices are compatible with classical quantitative chemistry.  Enzyme-kinetic observables are naturally connected with Michaelis--Menten and Briggs--Haldane type descriptions \cite{MichaelisMenten1913Invertinwirkung,BriggsHaldane1925EnzymeKinetics,JohnsonGoody2011OriginalMichaelisConstant,Bisswanger2017EnzymeKinetics}.  Thermodynamic observables are naturally connected with free-energy, heat, and equilibrium descriptions \cite{Alberty2003ThermodynamicsBiochemicalReactions,DillBromberg2010MolecularDrivingForces}.  Regulatory or cooperative observables may be connected with binding response, saturation fraction, or activity level \cite{Hill1910Haemoglobin,MonodWymanChangeux1965AllostericTransitions,KoshlandNemethyFilmer1966BindingModels}.

\subsection{Observable defect indices}

\begin{definition}[Observable distributivity defect]
\[
\delta_D^{\OO}(a;b,c;\gamma)=
\left|\OO(a\gamma(b+c))-\OO(a\gamma b)-\OO(a\gamma c)\right|.
\]
\end{definition}

The quantity $\delta_D^{\OO}(a;b,c;\gamma)$ measures the observable failure of right distributivity for the indicated states and mediator.  If
\[
\delta_D^{\OO}(a;b,c;\gamma)=0,
\]
then the aggregated input $b+c$ and the separated inputs $b,c$ give the same observable response under the mediator $\gamma$.  This may be interpreted as observable non-interference between the two channels.  If the defect is positive, then the observable response of the combined input differs from the sum of the separate responses.

A left-distributivity defect may be defined analogously by
\[
{}^{L}\delta_D^{\OO}(a,b;c;\gamma)=
\left|\OO((a+b)\gamma c)-\OO(a\gamma c)-\OO(b\gamma c)\right|.
\]
We focus mainly on the right version for notational simplicity; all interpretations have left-sided analogues.

\begin{definition}[Observable associativity defect]
\[
\delta_A^{\OO}(a,b,c;\gamma,\mu)=
\left|\OO((a\gamma b)\mu c)-\OO(a\gamma(b\mu c))\right|.
\]
\end{definition}

The quantity $\delta_A^{\OO}(a,b,c;\gamma,\mu)$ measures the observable failure of associativity for the two parenthesized mediated routes.  If
\[
\delta_A^{\OO}(a,b,c;\gamma,\mu)=0,
\]
then the two routes are indistinguishable through the observable $\OO$.  If the defect is positive, then the order in which the mediated interaction is grouped affects the observed result.  This provides a diagnostic for path-dependence, intermediate sensitivity, or kinetic memory.

In exact equilibrium-like regimes, one expects many relevant path-defects to vanish.  In contrast, sequential kinetic mechanisms, inhibited pathways, cooperative transitions, and regulatory mechanisms may produce nonzero defects.  Later sections use these defect indices to distinguish exact TGS regimes from regimes that are only approximately TGS or genuinely non-TGS at the operational level.

\subsection{Observable ratio indices}

Defect indices measure absolute deviation.  Ratio indices measure relative deviation.  Both are useful because an absolute defect may be small in magnitude but large relative to the baseline response, or large in magnitude but small relative to a high-output regime.

\begin{definition}[Distributivity ratio]
When $\OO(a\gamma b)+\OO(a\gamma c)>0$, define
\[
D_{\OO}(a;b,c;\gamma)=
\frac{\OO(a\gamma(b+c))}{\OO(a\gamma b)+\OO(a\gamma c)}.
\]
\end{definition}

The ratio $D_{\OO}$ compares the observable response of the combined channel with the sum of the observable responses of the separate channels.  The denominator condition excludes the zero-baseline case, where the ratio would be undefined.  In such a case, the defect $\delta_D^{\OO}$ may still be meaningful, but the ratio is not.

\begin{definition}[Associativity ratio]
When $\OO(a\gamma(b\mu c))>0$, define
\[
A_{\OO}(a,b,c;\gamma,\mu)=
\frac{\OO((a\gamma b)\mu c)}{\OO(a\gamma(b\mu c))}.
\]
\end{definition}

The ratio $A_{\OO}$ compares the observable outcome of the left-parenthesized mediated route with that of the right-parenthesized mediated route.  Again, the positivity condition in the denominator is required for the ratio to be defined.  When the denominator vanishes, the defect $\delta_A^{\OO}$ remains the safer diagnostic.

\subsection{Interpretation rules}

The following interpretation rules will be used throughout the paper.

First, if
\[
D_{\OO}(a;b,c;\gamma)=1,
\]
then the mediated response is observably additive across the two channels $b$ and $c$.  This is interpreted as observable non-interference.  If
\[
D_{\OO}(a;b,c;\gamma)<1,
\]
then the combined response is smaller than the sum of the separate responses.  This may indicate inhibition, competition for a mediator, saturation, crowding, or another suppressive effect.  If
\[
D_{\OO}(a;b,c;\gamma)>1,
\]
then the combined response exceeds the sum of the separate responses.  This may indicate synergy, cooperative activation, or amplification.  These interpretations are qualitative and must be tied to the chosen observable and modelling context; they are not automatic mechanistic conclusions.

Second, if
\[
A_{\OO}(a,b,c;\gamma,\mu)=1,
\]
then the two parenthesized mediated routes are observably path-independent.  If
\[
A_{\OO}(a,b,c;\gamma,\mu)\ne 1,
\]
then the routes are observably path-dependent.  Such path-dependence may arise from intermediate sensitivity, sequential kinetic constraints, regulatory memory, or non-equilibrium ordering effects.  Michaelis--Menten type kinetics and later refinements provide standard examples where intermediate formation and steady-state assumptions matter for the observed rate law \cite{MichaelisMenten1913Invertinwirkung,BriggsHaldane1925EnzymeKinetics,SchnellMendoza1997ClosedForm,CornishBowden2015OneHundredYears}.

Third, observable equality is weaker than algebraic equality.  The implication
\[
(a\gamma b)\mu c=a\gamma(b\mu c)
\quad\Longrightarrow\quad
\OO((a\gamma b)\mu c)=\OO(a\gamma(b\mu c))
\]
holds for every observable $\OO$.  However, the converse need not hold.  Even if
\[
\OO((a\gamma b)\mu c)=\OO(a\gamma(b\mu c)),
\]
the underlying states may still differ.  The converse can be asserted only under a separation assumption, for example if the chosen family of observables separates the relevant states.

Similarly, an observable distributivity defect equal to zero does not by itself prove algebraic distributivity in $S$.  It proves only that the selected observable cannot distinguish the two sides.  Therefore, all claims in this paper are stated carefully at either the algebraic level or the observable level.  This distinction prevents the misuse of numerical diagnostics as if they were automatically structural identities.
\section{Distributivity and Non-Interference}

This section gives the first operational interpretation of the compressed TGS laws.  Right distributivity is interpreted as an ideal non-interference principle for two parallel mediated channels.  The point is not that all chemical systems satisfy distributivity.  Rather, distributivity describes a special regime in which a mediated response to a combined input agrees with the sum of the mediated responses to the separate inputs.  When this equality fails at the observable level, the failure is interpreted as observable coupling, inhibition, competition, or synergy, depending on the chosen observable and chemical context.

\subsection{Two operational protocols}

Fix states $a,b,c\in S$ and a mediator $\gamma\in\Gamma$.  There are two natural operational protocols.

\subsubsection{Mixture protocol}

The first protocol combines $b$ and $c$ before applying the mediated operation:
\[
a\gamma(b+c).
\]
Chemically, this represents a situation in which the input channels $b$ and $c$ are presented together to the state $a$ under the same mediator $\gamma$.  The mediator may be a catalyst, solvent, inhibitor, thermodynamic condition, field, or regulatory context.

\subsubsection{Separate-protocol sum}

The second protocol applies the mediated operation separately and then aggregates the two outputs:
\[
a\gamma b+a\gamma c.
\]
Chemically, this represents two isolated mediated responses: first the response of $a$ to $b$ under $\gamma$, and second the response of $a$ to $c$ under the same $\gamma$.  The two outputs are then combined through the semigroup operation $+$.

\subsection{Chemical meaning of distributivity}

Right distributivity asserts that the mixture protocol and the separate-protocol sum agree:
\[
a\gamma(b+c)=a\gamma b+a\gamma c.
\]
Thus, the response to the combined input $b+c$ decomposes additively into the response to $b$ and the response to $c$.  Operationally, this is an ideal non-interference condition.  It says that the two channels represented by $b$ and $c$ do not compete for the mediator, do not inhibit one another, do not amplify one another, and do not create a new coupled output beyond the sum of their separate responses.

This interpretation is consistent with the role of distributive laws in semiring theory, where addition often represents parallel composition and multiplication-like operations represent structured transformation or sequential composition \cite{Golan1999SemiringsApplications,HebischWeinert1998Semirings,Tropashko2006TernaryRelationsSemiring}.  In the chemical setting, however, distributivity should be treated as an operational regime rather than as an automatic chemical truth.  Catalytic saturation, inhibitory competition, cooperative binding, or regulatory interaction may all produce observable deviations from distributivity \cite{BriggsHaldane1925EnzymeKinetics,MonodWymanChangeux1965AllostericTransitions,KoshlandNemethyFilmer1966BindingModels}.

\subsection{Proposition A: Observable non-interference}

\begin{proposition}[Distributivity implies observable non-interference]
Assume right distributivity holds for $a,b,c\in S$ and $\gamma\in\Gamma$. Assume also that $\OO$ is additive on the relevant output sum, i.e.
\[
\OO(a\gamma b+a\gamma c)=\OO(a\gamma b)+\OO(a\gamma c).
\]
Then
\[
\delta_D^{\OO}(a;b,c;\gamma)=0.
\]
If $\OO(a\gamma b)+\OO(a\gamma c)>0$, then
\[
D_{\OO}(a;b,c;\gamma)=1.
\]
\end{proposition}

\begin{proof}
By right distributivity,
\[
a\gamma(b+c)=a\gamma b+a\gamma c.
\]
Applying the observable $\OO$ gives
\[
\OO(a\gamma(b+c))
=
\OO(a\gamma b+a\gamma c).
\]
By the assumed additivity of $\OO$ on this particular output sum,
\[
\OO(a\gamma b+a\gamma c)
=
\OO(a\gamma b)+\OO(a\gamma c).
\]
Therefore
\[
\OO(a\gamma(b+c))
=
\OO(a\gamma b)+\OO(a\gamma c).
\]
Substituting this equality into the definition of the observable distributivity defect gives
\[
\delta_D^{\OO}(a;b,c;\gamma)
=
\left|
\OO(a\gamma(b+c))-\OO(a\gamma b)-\OO(a\gamma c)
\right|
=0.
\]

If, in addition,
\[
\OO(a\gamma b)+\OO(a\gamma c)>0,
\]
then the distributivity ratio is defined and satisfies
\[
D_{\OO}(a;b,c;\gamma)
=
\frac{\OO(a\gamma(b+c))}
{\OO(a\gamma b)+\OO(a\gamma c)}
=
\frac{\OO(a\gamma b)+\OO(a\gamma c)}
{\OO(a\gamma b)+\OO(a\gamma c)}
=1.
\]
This proves the proposition.
\end{proof}

The proposition separates two assumptions.  The first is algebraic distributivity in the TGS system.  The second is additivity of the observable on the relevant output sum.  Both are needed because an observable need not preserve the semigroup operation $+$ automatically.  For example, product concentration, absorbance, heat release, or fluorescence intensity may be additive only under specific calibration or non-interaction assumptions.  Therefore, observable non-interference follows safely only when the relevant algebraic and observable additivity hypotheses are both stated.

\subsection{Proposition B: Nonzero distributivity defect detects deviation from the separate-response prediction}

\begin{proposition}[Nonzero distributivity defect detects deviation from the separate-response prediction]
If
\[
\delta_D^{\OO}(a;b,c;\gamma)>0,
\]
then
\[
\OO(a\gamma(b+c))
\ne
\OO(a\gamma b)+\OO(a\gamma c).
\]
Thus the mixture protocol is distinguishable, under $\OO$, from the numerical separate-response prediction.  If, in addition, $\OO$ is additive on the relevant output sum, then it is distinguishable from the observable value of the separate-protocol sum $a\gamma b+a\gamma c$.
\end{proposition}

\begin{proof}
By definition,
\[
\delta_D^{\OO}(a;b,c;\gamma)=
\left|
\OO(a\gamma(b+c))-\OO(a\gamma b)-\OO(a\gamma c)
\right|.
\]
If this number is strictly positive, then
\[
\OO(a\gamma(b+c))
\ne
\OO(a\gamma b)+\OO(a\gamma c).
\]
This proves that the observable value of the mixture protocol differs from the numerical sum of the two separate responses.

If $\OO$ is also additive on the relevant output sum, then
\[
\OO(a\gamma b+a\gamma c)=\OO(a\gamma b)+\OO(a\gamma c).
\]
Combining this equality with the strict inequality above gives
\[
\OO(a\gamma(b+c))
\ne
\OO(a\gamma b+a\gamma c),
\]
so the mixture protocol and the algebraic separate-protocol sum are distinguishable by $\OO$ under that additional additivity hypothesis.
\end{proof}

The proposition gives a safe diagnostic.  A positive defect shows that the chosen observable detects a discrepancy between the mixture protocol and the numerical separate-response prediction.  It does not identify the physical cause by itself.  In different chemical contexts, the same formal sign of deviation may correspond to inhibition, competition, saturation, cooperative enhancement, or regulatory coupling.

\subsection{Separation caveat}

Observable diagnostics must be distinguished from algebraic identities.  The following conditional proposition records the safe direction in which observable inequality can be upgraded to state-level inequality.

\begin{proposition}[Separation caveat for distributivity]
Let
\[
x=a\gamma(b+c),
\qquad
y=a\gamma b+a\gamma c.
\]
Assume that the observable $\OO$ separates the two states $x$ and $y$ in the following local sense:
\[
\OO(x)\ne \OO(y)\quad\Longrightarrow\quad x\ne y.
\]
If
\[
\OO(a\gamma(b+c))\ne \OO(a\gamma b+a\gamma c),
\]
then
\[
a\gamma(b+c)\ne a\gamma b+a\gamma c.
\]
\end{proposition}

\begin{proof}
This is immediate from the stated local separation property.  Taking
\[
x=a\gamma(b+c)
\quad\text{and}\quad
y=a\gamma b+a\gamma c,
\]
the observable inequality gives $x\ne y$.  Hence right distributivity fails for the indicated states and mediator.
\end{proof}

The proposition is intentionally modest.  It does not claim that every observable separates all states.  In many chemical settings, two distinct coarse-grained states may have the same measured concentration, the same absorbance, or the same heat release.  Therefore, a zero observable defect does not prove algebraic distributivity, and a nonzero numerical defect proves algebraic non-distributivity only when the observable is known to separate the relevant outputs.

A stronger version may be stated for a family of observables.  If a family
\[
\mathcal{O}=\{\OO_i:S\to\mathbb{R}_{\geq 0}\}_{i\in I}
\]
separates the relevant states, then equality of all observables in the family can be used as a more reliable proxy for equality in $S$.  This is often closer to experimental practice, where one combines several measurements rather than relying on a single scalar response.

\subsection{Chemical interpretation}

A nonzero observable distributivity defect indicates that the response of the mixture protocol is not captured by adding the separate responses.  If
\[
D_{\OO}(a;b,c;\gamma)<1,
\]
then the combined response is smaller than the separate-protocol sum.  This may occur in inhibitory, competitive, saturation-limited, or resource-limited regimes.  For example, two substrates or channels may compete for the same catalytic site, or an inhibitor-like component may suppress the effect of another channel.  Such interpretations are compatible with standard kinetic reasoning in enzyme systems, where observable rates depend on substrate, enzyme, inhibitor, and intermediate structure rather than on purely additive response rules \cite{MichaelisMenten1913Invertinwirkung,BriggsHaldane1925EnzymeKinetics,RubinowLebowitz1970TimeDependentMichaelisMenten,Bisswanger2017EnzymeKinetics}.

If
\[
D_{\OO}(a;b,c;\gamma)>1,
\]
then the combined response is larger than the separate-protocol sum.  This may indicate synergy, cooperative activation, or amplification.  Cooperative binding and allosteric models provide classical motivation for regimes where the response of a combined system is not reducible to the sum of independent contributions \cite{Hill1910Haemoglobin,MonodWymanChangeux1965AllostericTransitions,KoshlandNemethyFilmer1966BindingModels}.  In the present TGS language, such behaviour is recorded as failure of observable non-interference.

If
\[
D_{\OO}(a;b,c;\gamma)=1
\]
and the relevant observable additivity assumptions are satisfied, then the two channels are observably non-interfering with respect to $\OO$.  This does not mean that the underlying microscopic mechanisms are independent in every possible sense.  It means only that, at the selected level of coarse-graining and measurement, the mixture protocol behaves like the sum of the separate protocols.

Thus, distributivity is best understood as an ideal operational regime.  It gives a clean algebraic expression for non-interference.  Defect and ratio indices then provide a way to describe departures from that ideal without forcing non-ideal chemical behaviour into an exact algebraic law.
\section{Associativity and Path-Independence}

This section gives the second operational interpretation of the compressed TGS laws.  Associativity is interpreted as path-independence of two bracketed mediated protocols.  The identity
\[
(a\gamma b)\mu c=a\gamma(b\mu c)
\]
does not mean that the two microscopic histories are necessarily identical.  It means that, at the chosen algebraic level of coarse-graining, the two ways of forming the mediated intermediate lead to the same final state.  When this equality fails after applying an observable, the failure is interpreted as observable path-dependence, kinetic memory, intermediate dependence, or order sensitivity.

\subsection{Two bracketed protocols}

Fix states $a,b,c\in S$ and mediators $\gamma,\mu\in\Gamma$.  There are two natural bracketed protocols.

\subsubsection{Left-associated protocol}

The first protocol forms the intermediate $a\gamma b$ before applying the second mediator:
\[
(a\gamma b)\mu c.
\]
Chemically, this represents a route in which the interaction between $a$ and $b$ under $\gamma$ is completed first, and the resulting intermediate then interacts with $c$ under $\mu$.

\subsubsection{Right-associated protocol}

The second protocol forms the intermediate $b\mu c$ before applying the first mediator:
\[
a\gamma(b\mu c).
\]
Chemically, this represents a route in which the interaction between $b$ and $c$ under $\mu$ is completed first, and the resulting intermediate then interacts with $a$ under $\gamma$.

\subsection{Chemical meaning of associativity}

Associativity asserts equality of the two bracketed protocols:
\[
(a\gamma b)\mu c=a\gamma(b\mu c).
\]
Thus, the final coarse-grained state is independent of the bracketing or order of intermediate formation.  In the present framework, associativity is therefore an algebraic expression of path-independence at the chosen level of coarse-graining.

This interpretation is deliberately modest.  It does not claim that the two physical histories are identical at every microscopic level.  Rather, it says that the state space $S$ is coarse enough, and the mediated operation is structured enough, that both bracketings lead to the same element of $S$.  In equilibrium-like regimes, such path-independence is natural.  In kinetic, inhibited, regulatory, or intermediate-sensitive regimes, one should expect associativity to fail either algebraically or observably.  Classical reaction-rate theory and enzyme kinetics show that intermediate formation and route structure may influence observed behaviour \cite{MichaelisMenten1913Invertinwirkung,BriggsHaldane1925EnzymeKinetics,Eyring1935ActivatedComplex,SchnellMendoza1997ClosedForm,CornishBowden2015OneHundredYears}.  The role of the present section is to isolate this distinction in TGS notation.

\subsection{Proposition C: Observable path-independence}

\begin{proposition}[Associativity implies observable path-independence]
If associativity holds for $a,b,c\in S$ and $\gamma,\mu\in\Gamma$, then for every observable $\OO:S\to\Rnn$,
\[
\delta_A^{\OO}(a,b,c;\gamma,\mu)=0.
\]
If $\OO(a\gamma(b\mu c))>0$, then
\[
A_{\OO}(a,b,c;\gamma,\mu)=1.
\]
\end{proposition}

\begin{proof}
By associativity,
\[
(a\gamma b)\mu c=a\gamma(b\mu c).
\]
Applying the observable $\OO$ to both sides gives
\[
\OO((a\gamma b)\mu c)=\OO(a\gamma(b\mu c)).
\]
Therefore
\[
\delta_A^{\OO}(a,b,c;\gamma,\mu)
=
\left|
\OO((a\gamma b)\mu c)-\OO(a\gamma(b\mu c))
\right|
=0.
\]

If, in addition,
\[
\OO(a\gamma(b\mu c))>0,
\]
then the associativity ratio is defined and satisfies
\[
A_{\OO}(a,b,c;\gamma,\mu)
=
\frac{\OO((a\gamma b)\mu c)}
{\OO(a\gamma(b\mu c))}
=
\frac{\OO(a\gamma(b\mu c))}
{\OO(a\gamma(b\mu c))}
=1.
\]
This proves the proposition.
\end{proof}

The proposition states the safe direction: algebraic associativity implies observable path-independence for every observable.  No additivity assumption on $\OO$ is needed here, because the comparison involves equality of two states rather than an algebraic sum of outputs.  This is why the associativity diagnostic is slightly cleaner than the distributivity diagnostic introduced in the previous section.

\subsection{Proposition D: Nonzero associativity defect detects kinetic memory}

\begin{proposition}[Nonzero associativity defect detects observable path-dependence]
If
\[
\delta_A^{\OO}(a,b,c;\gamma,\mu)>0,
\]
then the two bracketed mediated protocols are distinguishable under $\OO$.
\end{proposition}

\begin{proof}
By definition,
\[
\delta_A^{\OO}(a,b,c;\gamma,\mu)
=
\left|
\OO((a\gamma b)\mu c)-\OO(a\gamma(b\mu c))
\right|.
\]
If this number is strictly positive, then
\[
\OO((a\gamma b)\mu c)\ne \OO(a\gamma(b\mu c)).
\]
Hence the left-associated protocol and the right-associated protocol are distinguishable by the observable $\OO$.  This proves observable path-dependence.

The conclusion is stated only at the observable level.  It does not require, and does not assert, a complete microscopic explanation of the difference.  It only says that the selected observable detects a difference between the two bracketed mediated routes.
\end{proof}

This proposition provides a diagnostic for kinetic memory or intermediate dependence.  If the measured or modelled value of the final state depends on whether the intermediate $a\gamma b$ or the intermediate $b\mu c$ is formed first, then the system is not path-independent with respect to the observable $\OO$.

\subsection{Separation caveat for associativity}

As in the distributivity case, observable equality and algebraic equality must be separated.  A nonzero observable defect is strong enough to imply that the two observed outputs differ.  A zero observable defect only says that the chosen observable cannot distinguish the two bracketed outputs.

\begin{proposition}[Observable separation for associativity]
Let
\[
x=(a\gamma b)\mu c,
\qquad
y=a\gamma(b\mu c).
\]
Assume that the observable $\OO$ separates the two relevant states in the local sense that
\[
\OO(x)\ne \OO(y)\quad\Longrightarrow\quad x\ne y.
\]
If
\[
\delta_A^{\OO}(a,b,c;\gamma,\mu)>0,
\]
then
\[
(a\gamma b)\mu c\ne a\gamma(b\mu c).
\]
\end{proposition}

\begin{proof}
The strict positivity of the defect gives
\[
\OO((a\gamma b)\mu c)\ne \OO(a\gamma(b\mu c)).
\]
By the stated local separation property, the corresponding states are distinct:
\[
(a\gamma b)\mu c\ne a\gamma(b\mu c).
\]
Thus associativity fails for the indicated states and mediators.
\end{proof}

The converse direction remains unsafe without additional assumptions.  It may happen that
\[
(a\gamma b)\mu c\ne a\gamma(b\mu c)
\]
but
\[
\OO((a\gamma b)\mu c)=\OO(a\gamma(b\mu c))
\]
for a particular observable $\OO$.  For example, two distinct coarse-grained states may produce the same product concentration, the same heat release, or the same absorbance at a selected wavelength.  Therefore, a family of observables is often more reliable than a single observable when one wants to infer state-level distinctions.

\subsection{Chemical interpretation}

A nonzero associativity defect
\[
\delta_A^{\OO}(a,b,c;\gamma,\mu)>0
\]
indicates that the observed outcome depends on the bracketed route.  In chemical language, this may correspond to path-dependence, kinetic memory, intermediate dependence, or order sensitivity.  The left-associated route forms $a\gamma b$ first, while the right-associated route forms $b\mu c$ first.  If those intermediates are not equivalent for the subsequent transformation, the final observable response may differ.

This interpretation is especially natural in kinetic settings.  In enzyme and catalytic mechanisms, the intermediate formed during one stage of a process may influence the effective rate, accessibility, or output of later stages.  Michaelis--Menten and Briggs--Haldane theories already emphasize the role of intermediate enzyme--substrate formation in determining kinetic behaviour \cite{MichaelisMenten1913Invertinwirkung,BriggsHaldane1925EnzymeKinetics}.  Later analyses of time-dependent enzyme kinetics further show that route and time dependence can matter before ideal limiting regimes are reached \cite{RubinowLebowitz1970TimeDependentMichaelisMenten,SchnellMendoza1997ClosedForm,CornishBowden2013OriginsEnzymeKinetics}.  In the TGS framework, such dependence is represented by failure of the associativity identity or by nonzero observable associativity defect.

In equilibrium-like regimes, by contrast, one may expect path-independent behaviour after sufficient relaxation or coarse-graining.  The present framework treats associativity as an algebraic analogue of path-independence phenomena familiar from equilibrium thermodynamics, where certain state functions depend only on the state and not on the path by which the state is reached \cite{Alberty2003ThermodynamicsBiochemicalReactions,DillBromberg2010MolecularDrivingForces}.  This analogy should not be overstated.  We do not identify TGS associativity directly with Hess's law or with any particular thermodynamic law.  Instead, we use associativity as a structural model for path-independence at the level of mediated algebraic operations.

Thus, the role of associativity in this paper is diagnostic.  When it holds algebraically, every observable sees path-independence.  When it fails observably, the system displays route sensitivity at the chosen level of measurement.  The defect and ratio indices introduced earlier provide a controlled way to record this failure without forcing kinetic or regulatory behaviour into an exact equilibrium-like identity.

\section{Separating Observable Families and Algebraic Recovery}

The previous sections emphasized a one-way implication: algebraic distributivity or associativity implies equality after applying any observable.  The converse is false for a single observable in general, because one observable may not distinguish all coarse-grained states.  This section records a stronger and more useful recovery principle.  If a family of observables separates the relevant outputs, then vanishing observable defects for the whole family recover the corresponding algebraic law.

This layer is important for operational interpretation.  A single measurement may hide state-level differences, whereas a sufficiently rich family of observables can distinguish them.  The theorems below formalize this point without assuming that every observable is complete.

\subsection{Separating families}

\begin{definition}[Separating family of observables]
A family of observables
\[
\mathfrak O=\{\OO_i:S\longrightarrow \Rnn\}_{i\in I}
\]
is said to separate a pair of states $x,y\in S$ if
\[
\OO_i(x)=\OO_i(y)\quad\text{for all }i\in I
\]
implies
\[
x=y.
\]
The family separates $S$ if it separates every pair of states in $S$.
\end{definition}

The definition is deliberately local.  For the recovery of one distributive or associative identity, it is enough that the family separates the two outputs being compared.  Global separation of all states is stronger than necessary.

\subsection{Recovery of distributivity}

\begin{theorem}[Observable recovery of distributivity]
Let $a,b,c\in S$ and $\gamma\in\Gamma$.  Suppose that
\[
\mathfrak O=\{\OO_i:S\longrightarrow \Rnn\}_{i\in I}
\]
separates the two states
\[
x=a\gamma(b+c),
\qquad
 y=a\gamma b+a\gamma c.
\]
Assume also that each $\OO_i$ is additive on the relevant output sum:
\[
\OO_i(a\gamma b+a\gamma c)=\OO_i(a\gamma b)+\OO_i(a\gamma c).
\]
If
\[
\delta_D^{\OO_i}(a;b,c;\gamma)=0
\qquad\text{for every }i\in I,
\]
then
\[
a\gamma(b+c)=a\gamma b+a\gamma c.
\]
\end{theorem}

\begin{proof}
For each $i\in I$, the vanishing condition gives
\[
\OO_i(a\gamma(b+c))=\OO_i(a\gamma b)+\OO_i(a\gamma c).
\]
By the assumed additivity of $\OO_i$ on the relevant output sum,
\[
\OO_i(a\gamma b)+\OO_i(a\gamma c)=\OO_i(a\gamma b+a\gamma c).
\]
Hence
\[
\OO_i(a\gamma(b+c))=\OO_i(a\gamma b+a\gamma c)
\]
for every $i\in I$.  Since the family $\mathfrak O$ separates the two states
\[
a\gamma(b+c)
\quad\text{and}\quad
 a\gamma b+a\gamma c,
\]
it follows that
\[
a\gamma(b+c)=a\gamma b+a\gamma c.
\]
\end{proof}

This theorem gives a precise answer to the limitation noted earlier.  A single zero defect is generally insufficient to prove algebraic distributivity.  However, vanishing defects across a separating observable family, together with observable additivity on the relevant output sums, do recover the distributive identity.

\subsection{Recovery of associativity}

\begin{theorem}[Observable recovery of associativity]
Let $a,b,c\in S$ and $\gamma,\mu\in\Gamma$.  Suppose that
\[
\mathfrak O=\{\OO_i:S\longrightarrow \Rnn\}_{i\in I}
\]
separates the two states
\[
x=(a\gamma b)\mu c,
\qquad
 y=a\gamma(b\mu c).
\]
If
\[
\delta_A^{\OO_i}(a,b,c;\gamma,\mu)=0
\qquad\text{for every }i\in I,
\]
then
\[
(a\gamma b)\mu c=a\gamma(b\mu c).
\]
\end{theorem}

\begin{proof}
For every $i\in I$, the condition
\[
\delta_A^{\OO_i}(a,b,c;\gamma,\mu)=0
\]
means
\[
\OO_i((a\gamma b)\mu c)=\OO_i(a\gamma(b\mu c)).
\]
Since this equality holds for every $i\in I$, and since the family $\mathfrak O$ separates the two relevant states, we conclude that
\[
(a\gamma b)\mu c=a\gamma(b\mu c).
\]
\end{proof}

Associativity recovery is cleaner than distributivity recovery because no additivity assumption is required.  The two sides being compared are states, and the observable family is applied directly to those states.

\subsection{Interpretive consequence}

The recovery theorems show that observable diagnostics are not merely descriptive.  Under separation hypotheses, they can reconstruct algebraic laws from observable data.  This produces a hierarchy of strength:
\[
\text{algebraic law}
\Longrightarrow
\text{all observable defects vanish}
\]
for arbitrary observables, while
\[
\text{all defects vanish for a separating family}
\Longrightarrow
\text{algebraic law}
\]
under the stated separation and additivity assumptions.  Thus the gap between observation and algebra can be closed when the measurement family is rich enough.

\section{Approximate Operational Regimes}

Exact associativity and exact distributivity are useful ideal laws, but real mediated systems may only approximately satisfy them at the observable level.  This section introduces approximate operational regimes and proves simple but useful ratio bounds.  The point is to make the ``approximately TGS'' language quantitative: small observable defects imply controlled deviation of the corresponding ratio indices from one, provided the denominator is bounded away from zero.

\subsection{Observable approximate distributivity}

\begin{definition}[Observable $\eta$-distributivity]
Let $\eta\geq 0$.  A mediated operation is called observably $\eta$-right-distributive on a subset $X\subseteq S$ with respect to an observable $\OO$ if
\[
\delta_D^{\OO}(a;b,c;\gamma)\leq \eta
\]
for all $a,b,c\in X$ and all relevant $\gamma\in\Gamma$.
\end{definition}

When $\eta=0$, this recovers observable right-distributivity with respect to $\OO$.  When $\eta>0$ is small relative to the scale of the observable response, the system behaves as an approximately non-interfering regime.

\begin{theorem}[Small distributivity defect gives approximate non-interference]
Assume that the mediated operation is observably $\eta$-right-distributive on $X\subseteq S$ with respect to $\OO$.  Then for all $a,b,c\in X$ and all relevant $\gamma\in\Gamma$,
\[
\left|
\OO(a\gamma(b+c))-\bigl(\OO(a\gamma b)+\OO(a\gamma c)\bigr)
\right|
\leq \eta.
\]
If in addition
\[
\OO(a\gamma b)+\OO(a\gamma c)\geq m>0,
\]
then
\[
\left|D_{\OO}(a;b,c;\gamma)-1\right|\leq \frac{\eta}{m}.
\]
\end{theorem}

\begin{proof}
The first inequality is exactly the definition of observable $\eta$-right-distributivity.  For the ratio estimate, write
\[
B=\OO(a\gamma b)+\OO(a\gamma c).
\]
Then $B\geq m>0$ and
\[
\left|D_{\OO}(a;b,c;\gamma)-1\right|
=\left|\frac{\OO(a\gamma(b+c))}{B}-1\right|
=\frac{\left|\OO(a\gamma(b+c))-B\right|}{B}
\leq \frac{\eta}{m}.
\]
\end{proof}

The estimate says that an absolute defect bound becomes a relative ratio bound when the separate-protocol output is not too small.  This prevents misleading ratio inflation in near-zero denominator regimes.

\subsection{Observable approximate associativity}

\begin{definition}[Observable $\eta$-associativity]
Let $\eta\geq 0$.  A mediated operation is called observably $\eta$-associative on a subset $X\subseteq S$ with respect to an observable $\OO$ if
\[
\delta_A^{\OO}(a,b,c;\gamma,\mu)\leq \eta
\]
for all $a,b,c\in X$ and all relevant $\gamma,\mu\in\Gamma$.
\end{definition}

When $\eta=0$, this recovers observable path-independence with respect to $\OO$.  Positive values of $\eta$ describe controlled path-dependence at the level of the selected observable.

\begin{theorem}[Small associativity defect gives approximate path-independence]
Assume that the mediated operation is observably $\eta$-associative on $X\subseteq S$ with respect to $\OO$.  Then for all $a,b,c\in X$ and all relevant $\gamma,\mu\in\Gamma$,
\[
\left|
\OO((a\gamma b)\mu c)-\OO(a\gamma(b\mu c))
\right|
\leq \eta.
\]
If in addition
\[
\OO(a\gamma(b\mu c))\geq m>0,
\]
then
\[
\left|A_{\OO}(a,b,c;\gamma,\mu)-1\right|\leq \frac{\eta}{m}.
\]
\end{theorem}

\begin{proof}
The first inequality is exactly the definition of observable $\eta$-associativity.  For the ratio estimate, write
\[
B=\OO(a\gamma(b\mu c)).
\]
Then $B\geq m>0$ and
\[
\left|A_{\OO}(a,b,c;\gamma,\mu)-1\right|
=\left|\frac{\OO((a\gamma b)\mu c)}{B}-1\right|
=\frac{\left|\OO((a\gamma b)\mu c)-B\right|}{B}
\leq \frac{\eta}{m}.
\]
\end{proof}

\subsection{Operational meaning}

Approximate regimes are useful because exact algebraic laws may be too rigid for kinetic, inhibited, or regulatory settings.  A small distributivity defect means that mixture response is close to the separate-protocol prediction for the chosen observable.  A small associativity defect means that two bracketed routes are nearly path-independent at the chosen observational scale.  These statements are not claims of exact algebraic structure; they are controlled quantitative diagnostics.

Together with the recovery theorems of the previous section, the approximate estimates provide a three-level diagnostic framework: exact algebraic laws, exact observable recovery under separating families, and approximate observable laws under small uniform defects.

\section{Potential-Generated Equilibrium Regimes}

The preceding sections treated associativity and distributivity as operational laws and introduced observable diagnostics for their failure.  We now describe one mathematically controlled situation in which associativity can be proved rather than merely assumed.  The setting is an equilibrium-type regime generated by a strictly convex potential.  The theorem below is not intended to cover arbitrary kinetic or regulatory chemistry.  It gives a sufficient condition for path-independence when both bracketed protocols impose the same final constraint set and the final state is selected by a unique minimization principle.

\subsection{Admissible state domain and potential}

Throughout this section, the state space is assumed to carry additional convex structure.  Let
\[
S\subseteq V
\]
be a convex admissible state domain inside a real vector space $V$.  Elements of $S$ are interpreted as admissible coarse-grained states.  The convexity assumption means that if $x,y\in S$ and $0\leq t\leq 1$, then
\[
tx+(1-t)y\in S.
\]
This is a modelling assumption appropriate for equilibrium-type coarse-grained variables such as composition vectors, concentration profiles, or normalized state descriptors.  It is not assumed in the general TGS framework.

Let
\[
G:S\longrightarrow \mathbb{R}
\]
be a strictly convex potential.  Thus, for all distinct $x,y\in S$ and all $0<t<1$,
\[
G(tx+(1-t)y)<tG(x)+(1-t)G(y).
\]
The function $G$ may be interpreted abstractly as an equilibrium-selecting potential.  In thermodynamic applications, such potentials are often related to free-energy type quantities, but the present theorem only requires strict convexity and unique minimization.  We do not need to identify $G$ with a particular physical free energy.

For each admissible mediated pair, let
\[
C(a,b)\subseteq S
\]
denote the constraint set associated with the interaction of $a$ and $b$ under the equilibrium mediator.  We assume that all such constraint sets used below are nonempty, closed, and convex, and that the corresponding minimization problems have unique minimizers.  These assumptions are deliberately explicit.  Without nonemptiness, a minimizer need not exist; without suitable closure or compactness-type hypotheses, the infimum may fail to be attained; and without strict convexity on a convex constraint set, uniqueness may fail.

The strict-convexity setting is a standard mathematical mechanism for obtaining unique equilibrium states.  It is compatible with equilibrium and thermodynamic variational viewpoints, where stable states are often selected by minimizing a suitable potential under constraints \cite{Alberty2003ThermodynamicsBiochemicalReactions,DillBromberg2010MolecularDrivingForces}.  In the present article, however, the role of convex minimization is purely structural: it supplies a proof-secure source of associativity under a common final constraint hypothesis.

\subsection{Equilibrium operation}

\begin{definition}[Potential-generated equilibrium operation]
For an equilibrium mediator $\epsilon$, define
\[
a\epsilon b=\operatorname*{arg\,min}\{G(x):x\in C(a,b)\},
\]
whenever the minimizer exists and is unique.
\end{definition}

Thus, the operation $a\epsilon b$ is not an arbitrary mediated output.  It is the unique admissible state in the constraint set $C(a,b)$ at which the potential $G$ is minimized.  The mediator $\epsilon$ denotes an equilibrium selection regime.  It should not be confused with a general kinetic mediator, inhibitor, field, or regulatory condition.  Other mediators may generate non-associative or non-distributive behaviour; the present section concerns only the special equilibrium mediator $\epsilon$.

The operation is therefore defined only on those pairs $(a,b)$ for which the corresponding minimization problem is meaningful and has a unique solution.  In particular, the definition is partial unless the required existence and uniqueness hypotheses hold for every pair in the state domain.  For the theorem below, we assume the operation is defined for all intermediate expressions that occur in the two bracketed protocols.

\subsection{Global constraint compatibility}

The key issue is not merely that each binary equilibrium operation is obtained by minimizing the same potential.  Associativity requires compatibility of the final constraints imposed by the two bracketed protocols.

For three states $a,b,c\in S$, the left-associated protocol
\[
(a\epsilon b)\epsilon c
\]
first forms the equilibrium state $a\epsilon b$ subject to $C(a,b)$ and then combines that intermediate with $c$.  The right-associated protocol
\[
a\epsilon(b\epsilon c)
\]
first forms the equilibrium state $b\epsilon c$ subject to $C(b,c)$ and then combines $a$ with that intermediate.

In general, these two procedures may impose different final admissible sets.  Therefore, even if both stages are equilibrium minimizations, associativity need not follow.  The crucial hypothesis is that both bracketings impose the same final/global constraint set.  We write this common set as
\[
C(a,b,c)\subseteq S.
\]
Operationally, this means that the final admissible states allowed after the three-state equilibrium process do not depend on whether the intermediate is formed from $(a,b)$ first or from $(b,c)$ first.

Under this hypothesis, the two bracketed protocols are not treated as two unrelated minimization problems.  They are two descriptions of the same final constrained equilibrium problem:
\[
\operatorname*{arg\,min}\{G(x):x\in C(a,b,c)\}.
\]
For the theorem below, the phrase ``both bracketings impose the same global constraint set'' means that the final left-associated and right-associated outputs are defined by this same constrained minimization problem.  Equivalently, the final admissible constraint generated by $(a\epsilon b,c)$ and the final admissible constraint generated by $(a,b\epsilon c)$ coincide with the same set $C(a,b,c)$.
The associativity theorem follows from uniqueness of this minimizer.

\subsection{Theorem E: Potential-generated associativity under common constraints}

\begin{theorem}[Potential-generated equilibrium associativity]
Assume that $S$ is a convex admissible state domain, $G:S\to\mathbb R$ is strictly convex, all relevant constraint sets are nonempty, closed, and convex, and the minimization problems below have unique minimizers. Suppose both bracketings of $a,b,c$ impose the same global constraint set $C(a,b,c)$. Then
\[
(a\epsilon b)\epsilon c=a\epsilon(b\epsilon c).
\]
\end{theorem}

\begin{proof}
By hypothesis, the left-associated protocol imposes the global constraint set $C(a,b,c)$.  Therefore its final output is the unique minimizer of $G$ on this common set:
\[
(a\epsilon b)\epsilon c
=
\operatorname*{arg\,min}\{G(x):x\in C(a,b,c)\}.
\]

Similarly, the right-associated protocol imposes the same global constraint set $C(a,b,c)$.  Hence its final output is also the unique minimizer of $G$ on that same set:
\[
a\epsilon(b\epsilon c)
=
\operatorname*{arg\,min}\{G(x):x\in C(a,b,c)\}.
\]

Since the minimizer of $G$ on $C(a,b,c)$ is unique, the two displayed outputs must be equal.  Therefore
\[
(a\epsilon b)\epsilon c=a\epsilon(b\epsilon c).
\]
This proves associativity for the indicated states under the equilibrium mediator $\epsilon$.
\end{proof}

The theorem shows that associativity can be obtained from three ingredients: a common potential, a common final constraint set, and uniqueness of the constrained minimizer.  Each hypothesis is essential to the proof.  If the two bracketings lead to different final constraint sets, the outputs may be different even though each is individually an equilibrium minimizer.  If the minimizer is not unique, the two bracketings may select different minimizers from the same constraint set.  If the potential does not control the final selection, there is no reason for the two bracketed outputs to agree.

\subsection{Thermodynamic language safety}

The theorem should be read as an algebraic analogue of path-independence under strict equilibrium hypotheses.  It should not be identified directly with Hess's law, nor should it be presented as a new thermodynamic law.  Hess-type path-independence concerns state functions and thermochemical quantities in a specific physical setting.  The present result is an abstract TGS statement: if both bracketed mediated protocols reduce to the same uniquely solvable constrained minimization problem, then their final algebraic states coincide.

This distinction is important for avoiding overclaiming.  The theorem does not say that every equilibrium chemical process is associative.  It says that an equilibrium-mediated TGS operation is associative when the admissible state domain is convex, the potential is strictly convex, the relevant constraint sets are well behaved, the minimizers are unique, and both bracketings impose the same global constraint set.  Under these assumptions, the final state depends only on the common constrained equilibrium problem and not on the bracketing used to describe intermediate formation.

The result also explains why associativity may fail outside equilibrium regimes.  In kinetic systems, the intermediate formed first may affect later accessibility or rate behaviour.  In inhibited or cooperative systems, the presence of one channel may modify the response of another.  In regulatory or allosteric regimes, the effective transformation law may depend on the route by which a state is reached \cite{MichaelisMenten1913Invertinwirkung,BriggsHaldane1925EnzymeKinetics,MonodWymanChangeux1965AllostericTransitions,KoshlandNemethyFilmer1966BindingModels}.  Such cases fall outside the strict hypotheses of Theorem~E and are better treated through the observable defect and ratio indices introduced earlier.
\section{Symbolic Intermediate Kinetics}

The preceding section gave a sufficient condition for associativity in an equilibrium-generated regime.  We now record a complementary failure example.  The example is deliberately symbolic.  It is not a quantitative ODE model, not a derivation of a rate law, and not a replacement for Michaelis--Menten or Briggs--Haldane kinetics.  Its purpose is narrower: to show that a coarse TGS representation of an intermediate-dependent pathway may naturally fail associativity.

\subsection{Purpose of the symbolic model}

Many chemical and biochemical transformations proceed through intermediate states.  In such cases, the output of a first mediated step may become the essential input for a later step.  If the first intermediate is not formed, the later output may not be reached.  This order sensitivity is precisely the kind of behaviour that associativity can fail to capture.

The symbolic model below isolates this point in the smallest possible TGS language.  It does not attempt to encode concentrations, rate constants, quasi-steady-state assumptions, reversible binding, or time evolution.  Those belong to quantitative enzyme kinetics and are treated in the classical literature \cite{MichaelisMenten1913Invertinwirkung,BriggsHaldane1925EnzymeKinetics,JohnsonGoody2011OriginalMichaelisConstant,SchnellMendoza1997ClosedForm,CornishBowden2015OneHundredYears,Bisswanger2017EnzymeKinetics}.  Here we only encode the logical fact that an intermediate-dependent route may distinguish two bracketings.

\subsection{Michaelis--Menten type pathway reminder}

The classical Michaelis--Menten type scheme may be written in the familiar form
\[
E+S\rightleftharpoons ES\to E+P.
\]
Here $E$ denotes enzyme, $S$ substrate, $ES$ the enzyme--substrate intermediate complex, and $P$ product.  The historical Michaelis--Menten paper and the Briggs--Haldane refinement are standard starting points for this pathway structure \cite{MichaelisMenten1913Invertinwirkung,BriggsHaldane1925EnzymeKinetics}.  Modern discussions emphasize that the Michaelis constant, intermediate formation, and time-dependent kinetic assumptions require careful interpretation \cite{JohnsonGoody2011OriginalMichaelisConstant,CornishBowden2013OriginsEnzymeKinetics,CornishBowden2015OneHundredYears}.

The present paper does not use the full quantitative theory.  We use only the qualitative feature that a substrate-like state may first pass through an intermediate-like state before reaching a product-like state.

\subsection{Symbolic states and mediators}

Let
\[
s_S,\quad s_I,\quad s_P,\quad s_0
\]
be states in $S$.  These symbols are interpreted as follows:
\[
s_S=\text{substrate-like state},\qquad
s_I=\text{intermediate-like state},
\]
\[
s_P=\text{product-like state},\qquad
s_0=\text{neutral or background state}.
\]
Let $\gamma,\mu\in\Gamma$ be two mediators.  The mediator $\gamma$ represents the symbolic formation of the intermediate, while $\mu$ represents the symbolic conversion of the intermediate into the product.  Assume that the mediated operation satisfies
\[
s_S\gamma s_0=s_I,\qquad s_I\mu s_0=s_P,\qquad s_0\mu s_0=s_0.
\]
Finally, assume that the intermediate and product states are distinct:
\[
s_I\neq s_P.
\]

The condition
\[
s_S\gamma s_0=s_I
\]
says that the substrate-like state reaches the intermediate-like state under the mediator $\gamma$ in the presence of the background state $s_0$.  The condition
\[
s_I\mu s_0=s_P
\]
says that the intermediate-like state reaches the product-like state under the mediator $\mu$.  The condition
\[
s_0\mu s_0=s_0
\]
says that the background state is inert under the $\mu$-mediated background interaction.

\subsection{Theorem F: Intermediate-dependent non-associativity}

\begin{theorem}[Symbolic intermediate kinetics may be non-associative]
Let $s_S,s_I,s_P,s_0\in S$ and let $\gamma,\mu\in\Gamma$ satisfy
\[
s_S\gamma s_0=s_I,\qquad s_I\mu s_0=s_P,\qquad s_0\mu s_0=s_0,
\]
and assume $s_I\neq s_P$. Then
\[
(s_S\gamma s_0)\mu s_0\neq s_S\gamma(s_0\mu s_0).
\]
\end{theorem}

\begin{proof}
Using the first defining relation,
\[
s_S\gamma s_0=s_I.
\]
Therefore the left-associated expression becomes
\[
(s_S\gamma s_0)\mu s_0=s_I\mu s_0.
\]
Using the second defining relation,
\[
s_I\mu s_0=s_P.
\]
Hence
\[
(s_S\gamma s_0)\mu s_0=s_P.
\]

On the other hand, using the third defining relation,
\[
s_0\mu s_0=s_0.
\]
Therefore the right-associated expression becomes
\[
s_S\gamma(s_0\mu s_0)=s_S\gamma s_0.
\]
Using the first defining relation again,
\[
s_S\gamma s_0=s_I.
\]
Hence
\[
s_S\gamma(s_0\mu s_0)=s_I.
\]

Thus
\[
(s_S\gamma s_0)\mu s_0=s_P
\qquad\text{and}\qquad
s_S\gamma(s_0\mu s_0)=s_I.
\]
Since $s_I\neq s_P$ by assumption, it follows that
\[
(s_S\gamma s_0)\mu s_0\neq s_S\gamma(s_0\mu s_0).
\]
This proves the theorem.
\end{proof}

\subsection{Interpretation}

Theorem~F shows that a coarse symbolic TGS representation of an intermediate-dependent Michaelis--Menten type pathway may be non-associative.  The reason is simple: the left-associated protocol forms the intermediate first and then converts it to the product, while the right-associated protocol applies the second mediator only to the background state before the substrate-like state is acted upon.  These two symbolic routes need not lead to the same coarse state.

The conclusion is intentionally limited.  We do not claim that the whole Michaelis--Menten theory is non-associative.  The full theory concerns concentrations, rate constants, reversible complex formation, assumptions such as rapid equilibrium or quasi-steady state, and time-dependent behaviour \cite{BriggsHaldane1925EnzymeKinetics,RubinowLebowitz1970TimeDependentMichaelisMenten,SchnellMendoza1997ClosedForm,Bisswanger2008EnzymeKinetics}.  The present example says only that if a TGS model records the intermediate as a distinct state and treats the formation and conversion steps as separate mediators, then the two bracketed algebraic protocols can differ.

In the language of the observable diagnostics introduced earlier, if an observable $\OO$ distinguishes the intermediate and product states, namely
\[
\OO(s_I)\neq \OO(s_P),
\]
then
\[
\delta_A^{\OO}(s_S,s_0,s_0;\gamma,\mu)
=
\left|\OO(s_P)-\OO(s_I)\right|>0.
\]
Thus the non-associativity becomes experimentally or computationally visible at the level of that observable.  If $\OO$ does not distinguish $s_I$ from $s_P$, then the algebraic non-associativity may be hidden from that particular observable.  This again illustrates the importance of separating algebraic equality from observable equality.
\section{Regulatory Coupling and Allosteric-Type\ Non-Distributivity}

This section discusses another common source of operational failure: non-distributivity caused by regulatory coupling.  The discussion is intentionally cautious.  We do not build a full quantitative allosteric model, nor do we derive regulatory rate laws.  Instead, we show how the observable distributivity defect introduced earlier can be used as a structural diagnostic for mixture effects that are not reducible to separate-channel responses.

\subsection{Cautious modelling setup}

Let $a\in S$ denote an enzyme-like, catalyst-like, or active chemical state.  Let $b\in S$ denote a substrate-like state, and let $c\in S$ denote an inhibitor-like, activator-like, or regulatory state.  Let $\gamma\in\Gamma$ be a catalytic or environmental mediator.  In this interpretation,
\[
a\gamma b
\]
represents the mediated response of the active state $a$ to the substrate-like state $b$, while
\[
a\gamma c
\]
represents the mediated response of $a$ to the regulatory state $c$.

The combined expression
\[
a\gamma(b+c)
\]
represents the response of the active state when the substrate-like and regulatory states are presented together.  The point is that the regulatory state $c$ may not behave as an independent additive input.  It may suppress, enhance, reshape, or redirect the response to $b$.  Classical allosteric and cooperative binding models show that regulatory effects can modify observed response curves in a non-additive way \cite{Hill1910Haemoglobin,MonodWymanChangeux1965AllostericTransitions,KoshlandNemethyFilmer1966BindingModels}.  The present TGS discussion uses these references only as conceptual motivation for non-independent channel behaviour.

\subsection{Protocol comparison}

The two protocols are
\[
a\gamma(b+c)
\qquad\text{versus}\qquad
a\gamma b+a\gamma c.
\]
The first is the regulatory mixture protocol.  It records the mediated response when the substrate-like and regulatory states are present together.  The second is the separate-protocol sum.  It records the response to $b$ alone and the response to $c$ alone, followed by formal aggregation of the two outputs.

If right distributivity holds, then
\[
a\gamma(b+c)=a\gamma b+a\gamma c.
\]
In that case, the regulatory state $c$ does not create an additional coupling effect at the algebraic level.  If right distributivity fails, then the response to the combined input is not captured by adding the separate responses.  In a regulatory interpretation, this failure is a natural algebraic signature of coupling between the substrate-like and regulatory channels.

\subsection{Observable non-distributivity}

At the observable level, non-distributivity is measured by
\[
\delta_D^{\OO}(a;b,c;\gamma)=
\left|
\OO(a\gamma(b+c))-\OO(a\gamma b)-\OO(a\gamma c)
\right|.
\]
If
\[
\delta_D^{\OO}(a;b,c;\gamma)>0,
\]
then the observable response of the mixture protocol differs from the sum of the observable responses of the separate protocols.  This is observable non-distributivity.

In a regulatory setting, such a positive defect may indicate inhibition, activation, regulatory coupling, competition for a binding site, cooperative enhancement, or allosteric-type behaviour.  The direction of the effect may be summarized by the ratio
\[
D_{\OO}(a;b,c;\gamma)=
\frac{\OO(a\gamma(b+c))}
{\OO(a\gamma b)+\OO(a\gamma c)}
\]
when the denominator is positive.  If
\[
D_{\OO}(a;b,c;\gamma)<1,
\]
the mixture response is smaller than the separate-protocol prediction, suggesting a suppressive or inhibitory effect.  If
\[
D_{\OO}(a;b,c;\gamma)>1,
\]
the mixture response is larger than the separate-protocol prediction, suggesting activation, amplification, or cooperative enhancement.  If
\[
D_{\OO}(a;b,c;\gamma)=1,
\]
the channels are observably non-interfering with respect to $\OO$.

These interpretations are qualitative.  A positive defect alone does not identify a molecular mechanism.  To claim a specific allosteric mechanism, one would need a detailed binding model, kinetic law, structural mechanism, or experimental validation.  The present framework supplies only a coarse algebraic diagnostic.

\subsection{Observable regulatory coupling}

\begin{proposition}[Observable regulatory coupling]
If $\delta_D^{\OO}(a;b,c;\gamma)>0$ in a regulatory interpretation of $a,b,c,\gamma$, then the regulatory mixture protocol and the separate-protocol sum are distinguishable by $\OO$.
\end{proposition}

\begin{proof}
By definition,
\[
\delta_D^{\OO}(a;b,c;\gamma)
=
\left|
\OO(a\gamma(b+c))-\OO(a\gamma b)-\OO(a\gamma c)
\right|.
\]
If this number is strictly positive, then
\[
\OO(a\gamma(b+c))
\ne
\OO(a\gamma b)+\OO(a\gamma c).
\]
Hence the observable response of the regulatory mixture protocol is not equal to the observable sum of the two separate responses.  Therefore the two protocols are distinguishable by $\OO$.
\end{proof}

This proposition is essentially the regulatory specialization of the earlier nonzero-defect theorem.  Its value is interpretive: it says that, in a regulatory modelling context, a positive distributivity defect is evidence of observable coupling between the substrate-like and regulatory inputs.

\subsection{Claim boundary}

The present section does not claim to model allostery completely.  Allosteric regulation involves conformational states, ligand binding, cooperative response, structural transitions, and kinetic or thermodynamic parameters.  Classical models such as the Hill description, the Monod--Wyman--Changeux model, and the Koshland--N{\'e}methy--Filmer model provide detailed mathematical frameworks for such phenomena \cite{Hill1910Haemoglobin,MonodWymanChangeux1965AllostericTransitions,KoshlandNemethyFilmer1966BindingModels}.  The TGS formulation here is more abstract and more modest.

The safe claim is the following.  If a regulatory or allosteric-type situation is represented at a coarse symbolic level by states $a,b,c$, a mediator $\gamma$, and an observable $\OO$, then failure of
\[
\OO(a\gamma(b+c))
=
\OO(a\gamma b)+\OO(a\gamma c)
\]
records observable non-distributivity.  This non-distributivity is consistent with regulatory coupling, but it is not by itself a complete mechanistic theory of regulation or allostery.

Thus, the TGS framework provides a diagnostic language for regulatory non-additivity.  It can identify where exact distributivity is inappropriate and where defect or ratio indices are more honest descriptions.  Any stronger biochemical claim must be supported by additional rate laws, binding equations, structural assumptions, or experimental data.
\section{Regime Classification}

The preceding sections separated two operational questions.  The first asks whether a mediated response to a combined input decomposes into the sum of separate mediated responses.  This is the distributivity question.  The second asks whether two bracketed mediated pathways lead to the same final coarse-grained state.  This is the associativity question.  Combining these two axes gives a useful structural classification of selected TGS operational regimes.

\subsection{Classification principle}

The classification is made at the chosen coarse-graining level.  Thus, a regime may be associative or distributive with respect to one state space, mediator set, and observable family, but fail to be so after a finer representation is introduced.  The table below should therefore be read as an operational guide, not as a universal taxonomy of chemistry.

The horizontal axis separates distributive and non-distributive regimes.  Distributive regimes satisfy an ideal non-interference law such as
\[
a\gamma(b+c)=a\gamma b+a\gamma c.
\]
Non-distributive regimes exhibit mixture effects, coupling, inhibition, activation, saturation, or other non-additive behaviour.

The vertical axis separates associative and non-associative regimes.  Associative regimes satisfy a path-independence law such as
\[
(a\gamma b)\mu c=a\gamma(b\mu c).
\]
Non-associative regimes exhibit order sensitivity, intermediate dependence, kinetic memory, or route-dependent behaviour.

This classification is structural and non-exhaustive.  It organizes the regimes naturally visible through the two main TGS laws studied in this paper.  It does not claim to classify all chemical mechanisms, nor does it replace detailed kinetic, thermodynamic, or regulatory models.

\subsection{Fourfold regime table}

\begin{center}
\renewcommand{\arraystretch}{1.35}
\begin{tabular}{|p{0.22\textwidth}|p{0.34\textwidth}|p{0.34\textwidth}|}
\hline
 & \textbf{Distributive} & \textbf{Non-distributive} \\
\hline
\textbf{Associative}
& Ideal equilibrium aggregation; ideal linear mediated channels.
& Path-independent systems with coupled or non-ideal mixture effects. \\
\hline
\textbf{Non-associative}
& Order-sensitive kinetic pathways with approximately independent channels.
& Regulatory, inhibited, cooperative, or strongly coupled kinetic systems. \\
\hline
\end{tabular}
\end{center}

\subsection{Reading the table}

The associative--distributive quadrant represents the most ideal TGS regime.  Here mediated pathways are path-independent, and parallel channels do not interfere with one another.  This is the natural location for equilibrium-like coarse-grained models with additive channel behaviour.  The potential-generated equilibrium construction of Section~8 gives one proof-secure mechanism for associativity under strict hypotheses.

The associative--non-distributive quadrant represents path-independent systems with non-ideal mixture behaviour.  In such a regime, different bracketings of a mediated pathway may lead to the same final coarse-grained state, while mixture protocols still fail to decompose additively.  This may occur when the final state is equilibrium-selected but the response to combined inputs involves saturation, competition, or other non-additive effects.

The non-associative--distributive quadrant represents order-sensitive pathways whose parallel channels remain approximately independent.  This is a plausible structural location for some kinetic or intermediate-dependent processes: the order of intermediate formation matters, but separate channels may still behave additively with respect to the selected observable.  The symbolic intermediate example in Section~9 illustrates how non-associativity can arise when the formation of an intermediate is essential for reaching a product-like state.

The non-associative--non-distributive quadrant represents the most strongly coupled regime.  Here both bracketing and mixture structure matter.  This quadrant includes coarse descriptions of regulatory, inhibited, cooperative, or strongly coupled kinetic systems, where route order and channel interaction both affect the observable outcome.  The regulatory discussion in Section~10 belongs naturally to this quadrant when the observable distributivity defect is positive and the system also displays path-dependence.

\subsection{Reviewer-safety note}

The table does not classify all chemical systems.  It organizes selected TGS operational regimes according to two structural axes: associativity and distributivity.  A real chemical system may require additional axes, such as reversibility, stochasticity, spatial dependence, time dependence, conservation constraints, thermodynamic admissibility, or multi-observable separation.  Therefore, the table should be used as a diagnostic map for the present algebraic framework, not as a complete or sharp chemical classification.
\section{Limitations and Scope}

This section records the boundaries of the framework.  The purpose is to prevent the operational TGS language from being read as a full physical, kinetic, thermodynamic, or biochemical theory.  The results in this paper are algebraic and structural.  They are intended to clarify when associativity and distributivity may be interpreted as path-independence and non-interference, and when their observable failures may be used as diagnostics.

\subsection{Algebraic and coarse-grained nature}

The framework is algebraic and coarse-grained.  The elements of $S$ are not assumed to be complete microscopic descriptions of chemical systems.  They may represent coarse states, symbolic states, phase labels, concentration classes, pathway states, or model-dependent descriptors.  Similarly, elements of $\Gamma$ are mediators in an abstract sense and may summarize catalysts, solvents, inhibitors, environmental conditions, fields, or regulatory contexts.

Therefore, a TGS-chemical system should not be interpreted as a direct molecular simulation model.  It does not automatically encode molecular geometry, reaction coordinates, stochastic fluctuations, spatial diffusion, quantum dynamics, or detailed rate constants.  Those features require additional modelling structure beyond the algebraic data
\[
(S,\Gamma,+,\circ).
\]
The value of the present framework lies in identifying structural regimes of mediated transformation, not in replacing detailed chemical models.

\subsection{Dependence on observables}

The defect and ratio indices depend on the chosen observable
\[
\OO:S\longrightarrow \mathbb{R}_{\geq 0}.
\]
Different observables may detect different aspects of the same coarse-grained state.  For example, product concentration, absorbance, heat release, fluorescence intensity, conversion ratio, and catalytic activity may lead to different diagnostics.  Thus, an observable defect is always relative to the observable used.

Moreover, observable equality does not imply algebraic equality unless suitable separation assumptions are stated.  It may happen that
\[
\OO(x)=\OO(y)
\]
while
\[
x\neq y
\]
inside $S$.  Hence
\[
\delta_A^{\OO}(a,b,c;\gamma,\mu)=0
\]
does not automatically prove
\[
(a\gamma b)\mu c=a\gamma(b\mu c),
\]
and
\[
\delta_D^{\OO}(a;b,c;\gamma)=0
\]
does not automatically prove
\[
a\gamma(b+c)=a\gamma b+a\gamma c.
\]
The reverse direction is safer: algebraic equality implies equality under every observable, but equality under one observable need not imply algebraic equality.  The recovery theorems in Section~6 require a separating family of observables; without that hypothesis they should not be used to infer algebraic identities from measured or computed equality.

\subsection{Symbolic kinetic example}

The Michaelis--Menten type example in Section~9 is symbolic.  It is not a substitute for the ordinary differential equation formulation of enzyme kinetics, nor does it derive the Michaelis--Menten rate law.  It does not encode enzyme concentration, substrate concentration, product concentration, rate constants, reversible binding constants, quasi-steady-state hypotheses, rapid-equilibrium assumptions, or transient dynamics.

Its purpose is only to show that an intermediate-dependent pathway can be represented in compressed TGS notation in a way that naturally fails associativity.  The conclusion is therefore limited to the symbolic level:
\[
(s_S\gamma s_0)\mu s_0\neq s_S\gamma(s_0\mu s_0)
\]
under the stated symbolic assumptions.  No claim is made that the full Michaelis--Menten theory is itself non-associative.

\subsection{Approximation bounds}

The approximate-regime estimates in Section~7 are relative to the chosen observable and to a positive denominator bound.  If the baseline denominator is close to zero, a small absolute defect may still produce an unstable or misleading ratio.  Thus, the estimates should be read together with the explicit lower-bound assumption \(m>0\).  They provide controlled approximate diagnostics, not experimental error analysis or statistical confidence intervals.

\subsection{Equilibrium theorem hypotheses}

The potential-generated equilibrium theorem depends on strict hypotheses.  It assumes a convex admissible state domain, a strictly convex potential, nonempty closed convex constraint sets, existence and uniqueness of all relevant minimizers, and a common global constraint set for the two bracketings.  These assumptions are not automatic in chemical systems.

In particular, associativity follows in Theorem~E because both bracketed protocols reduce to the same uniquely solvable constrained minimization problem:
\[
\operatorname*{arg\,min}\{G(x):x\in C(a,b,c)\}.
\]
If the two bracketings impose different global constraint sets, if minimizers are not unique, or if the potential does not control the final state selection, then the proof no longer applies.  Thus, the theorem gives a sufficient condition for equilibrium-type associativity, not a general law for all mediated chemical transformations.

\subsection{Regime table boundary}

The regime table is structural and non-exhaustive.  It organizes selected TGS operational regimes using two axes: associativity and distributivity.  These axes are useful for distinguishing path-independence, path-dependence, non-interference, and regulatory or mixture coupling.  However, they do not exhaust the features needed to classify real chemical systems.

A more detailed classification may require additional axes, including reversibility, time dependence, stochasticity, thermodynamic admissibility, conservation laws, spatial transport, multi-scale structure, or empirical identifiability.  Therefore, the table should be read as a diagnostic map for the present algebraic framework, not as a complete chemical taxonomy.

\subsection{Citation and validation boundary}

The citations used in this paper support mathematical and chemical background.  They justify the use of semiring, ternary, kinetic, thermodynamic, and allosteric language, and they situate the present model relative to established literature.  They do not by themselves provide experimental validation of the TGS framework.

No experimental data are analysed in this paper.  No new kinetic constants, binding parameters, thermodynamic values, or reaction mechanisms are inferred.  Unless a theorem is explicitly derived from stated assumptions, citations should be read as background support rather than as proof of the paper's algebraic claims.  The mathematical claims of the paper rest on the definitions, hypotheses, and proofs given here.
\section{Conclusion}

\subsection{Summary of proved content}

This paper studied the operational meaning of associativity and distributivity in compressed TGS-chemical systems.  The focus was not on giving a second axiomatization of TGS-chemical systems, but on analysing the consequences and failure regimes of the algebraic laws introduced in the foundational MATCH article \cite{GokavarapuMATCH2026Axiomatic}.

The first proved result was that right distributivity implies observable non-interference, provided the chosen observable is additive on the relevant output sum.  More precisely, if
\[
a\gamma(b+c)=a\gamma b+a\gamma c
\]
and
\[
\OO(a\gamma b+a\gamma c)=\OO(a\gamma b)+\OO(a\gamma c),
\]
then
\[
\delta_D^{\OO}(a;b,c;\gamma)=0.
\]
When the denominator is positive, this also gives
\[
D_{\OO}(a;b,c;\gamma)=1.
\]
The second result showed that a nonzero distributivity defect detects observable distinguishability between the mixture protocol and the separate-protocol sum.

The third proved result was that associativity implies observable path-independence for every observable.  If
\[
(a\gamma b)\mu c=a\gamma(b\mu c),
\]
then
\[
\delta_A^{\OO}(a,b,c;\gamma,\mu)=0
\]
for every observable $\OO:S\to\mathbb{R}_{\geq 0}$.  When the denominator is positive, this also gives
\[
A_{\OO}(a,b,c;\gamma,\mu)=1.
\]
The fourth result showed that a nonzero associativity defect detects observable distinguishability between the two bracketed mediated protocols.

The next theorem layer established recovery results for separating observable families.  In the distributive case, if a family of observables separates the two states
\[
a\gamma(b+c)
\quad\text{and}\quad
 a\gamma b+a\gamma c,
\]
and all corresponding observable distributivity defects vanish under the stated additivity assumptions, then the algebraic distributive identity follows.  In the associative case, if a family of observables separates
\[
(a\gamma b)\mu c
\quad\text{and}\quad
 a\gamma(b\mu c),
\]
and all corresponding observable associativity defects vanish, then algebraic associativity follows for the indicated data.

The approximate-regime layer proved that uniform small defect bounds give controlled ratio estimates.  If the distributivity defect is bounded by \(\eta\) and the separate-protocol denominator is at least \(m>0\), then
\[
\left|D_{\OO}(a;b,c;\gamma)-1\right|\leq \frac{\eta}{m}.
\]
Similarly, if the associativity defect is bounded by \(\eta\) and the right-associated denominator is at least \(m>0\), then
\[
\left|A_{\OO}(a,b,c;\gamma,\mu)-1\right|\leq \frac{\eta}{m}.
\]

The following main structural result was the potential-generated equilibrium theorem.  Under strict hypotheses---convex admissible state domain, strictly convex potential, nonempty closed convex constraint sets, unique minimizers, and a common global constraint set for both bracketings---the equilibrium operation satisfies
\[
(a\epsilon b)\epsilon c=a\epsilon(b\epsilon c).
\]
Thus, associativity follows when both bracketed protocols reduce to the same uniquely solvable constrained minimization problem.

Finally, the paper constructed a symbolic intermediate-dependent Michaelis--Menten type example.  Under the assumptions
\[
s_S\gamma s_0=s_I,\qquad s_I\mu s_0=s_P,\qquad s_0\mu s_0=s_0,
\]
with
\[
s_I\neq s_P,
\]
one obtains
\[
(s_S\gamma s_0)\mu s_0\neq s_S\gamma(s_0\mu s_0).
\]
This gives a concrete symbolic example of non-associativity caused by intermediate dependence.

\subsection{Chemical meaning}

The results support a structural distinction between equilibrium-like and kinetic or regulatory regimes.  Equilibrium-type regimes may satisfy path-independence when a common potential and common final constraint set determine the final state.  Kinetic or intermediate-dependent regimes may fail associativity because the order of intermediate formation matters.  Regulatory, inhibited, cooperative, or allosteric-type regimes may fail distributivity because the response to a mixture is not the sum of the separate responses.

The observable defect and ratio indices provide a bridge between algebraic laws and possible measurements.  The added recovery and approximation results strengthen this bridge by identifying when observable data recover algebraic laws and when small defects give controlled approximate laws.  A distributivity defect measures the observable failure of non-interference.  An associativity defect measures the observable failure of path-independence.  The corresponding ratios distinguish ideal behaviour from suppressive, amplifying, or route-sensitive behaviour, subject to the chosen observable and denominator conditions.

This connection is intentionally modest.  A nonzero defect indicates observable distinguishability, not a complete mechanistic explanation.  Likewise, a zero defect for one observable does not prove algebraic equality unless suitable separation assumptions are available.  The paper therefore keeps algebraic equality, observable equality, and chemical interpretation as distinct layers.

\subsection{Future work}

Several directions remain open.  One direction is to develop quantitative kinetic TGS models in which the mediated operation carries explicit rate parameters, time dependence, or concentration dependence.  Such models would allow the symbolic intermediate example to be replaced by a genuine kinetic construction.

A second direction is to study stochastic TGS models.  In such a setting, the output of a mediated operation may be a distribution over states rather than a single state.  Associativity and distributivity defects could then be measured using expected observables, variances, or divergence-type quantities.

A third direction is network-level TGS chemistry.  Instead of studying individual triples of states and mediators, one may study reaction networks whose nodes are TGS states and whose mediated edges or hyperedges encode parameter-dependent transformations.  This could connect the present framework with network chemistry and symbolic computational models.

A fourth direction is data-driven estimation of observable defects.  Given experimental or simulated observations, one could estimate quantities such as
\[
\delta_D^{\OO}(a;b,c;\gamma)
\qquad\text{and}\qquad
\delta_A^{\OO}(a,b,c;\gamma,\mu)
\]
to identify where non-interference or path-independence is a reasonable approximation and where it fails.

\subsection{Final caution}

The paper does not claim to give a complete theory of chemical transformations.  It does not claim that all chemical systems are TGS systems, that all equilibrium systems are associative, or that all kinetic systems are non-associative.  It also does not claim experimental verification of the proposed framework.

The safe conclusion is narrower.  Within the compressed TGS setting, distributivity can be read as observable non-interference under stated additivity assumptions, associativity can be read as path-independence, and the failure of these laws can be measured by observable defect and ratio indices.  Under strict equilibrium hypotheses, associativity can be proved.  Under a symbolic intermediate-dependent construction, associativity can fail.  These results provide an operational extension of the published TGS-chemical foundation and a cautious basis for further mathematical development.

\section*{Acknowledgements}
The authors thank the Commissioner of Higher Education, Government of Andhra Pradesh, India, and the Principal, Government College (Autonomous), Rajahmundry, for providing a supportive academic environment.  The authors also thank the Department of Mathematics, Acharya Nagarjuna University, Guntur, Andhra Pradesh.  The authors further thank colleagues in the Departments of Mathematics and Chemistry for useful discussions and encouragement during the preparation of this work.

\section*{Declarations}
\noindent\textbf{Funding.} No funding was received for this work.

\noindent\textbf{Conflict of interest.} The authors declare no conflict of interest.

\noindent\textbf{Data availability.} No datasets were generated or analysed during the preparation of this article.

\noindent\textbf{Ethics approval.} Not applicable.

\noindent\textbf{Author contribution.} Chandrasekhar Gokavarapu coordinated the manuscript and prepared the primary mathematical draft.  Venkata Rao Kaviti and Srinivas Thirunagari contributed to the chemical interpretation of mediated transformations and regulatory examples.  Sajani Lavanya Madasi contributed to algebraic organization, verification of definitions, and manuscript revision.  Madhusudhana Rao Dasari contributed to conceptual supervision, mathematical review, and final revision.  All authors reviewed and approved the final manuscript.

\appendix
\section{Additional Symbolic Examples}

This appendix gives four explicit toy examples realizing the four operational regimes in the classification table.  The examples are purely algebraic.  They are not intended as quantitative chemical models.  Their purpose is to show that the four combinations of associativity and distributivity are logically distinct.

Throughout this appendix, take the mediator set to be a singleton
\[
\Gamma=\{\gamma\}.
\]
We write
\[
a*b
\]
instead of
\[
a\gamma b
\]
to simplify notation.

\subsection{Example 1: Distributive and associative toy model}

Let
\[
S=\mathbb{R}_{\geq 0}
\]
with the usual addition.  Define
\[
a*b=ab
\]
for all $a,b\in S$.

\subsubsection*{Associativity check}

For all $a,b,c\in S$,
\[
(a*b)*c=(ab)c=abc,
\]
while
\[
a*(b*c)=a(bc)=abc.
\]
Hence
\[
(a*b)*c=a*(b*c).
\]
Thus the operation is associative.

\subsubsection*{Distributivity check}

For all $a,b,c\in S$,
\[
a*(b+c)=a(b+c)=ab+ac=a*b+a*c.
\]
Similarly,
\[
(a+b)*c=(a+b)c=ac+bc=a*c+b*c.
\]
Hence the operation is both right and left distributive over $+$.

Therefore this example lies in the associative--distributive quadrant.  It represents the ideal regime in which path-independence and non-interference both hold.

\subsection{Example 2: Distributive but non-associative toy model}

Let
\[
S=\mathbb{N}^2
\]
with componentwise addition:
\[
(x_1,x_2)+(y_1,y_2)=(x_1+y_1,x_2+y_2).
\]
Let
\[
e_1=(1,0),\qquad e_2=(0,1).
\]
Define a bilinear operation $*$ on the basis vectors by
\[
e_1*e_1=e_2,\qquad e_2*e_1=e_1,
\]
and
\[
e_1*e_2=0,\qquad e_2*e_2=0,
\]
where
\[
0=(0,0).
\]
Extend this operation bilinearly over $\mathbb{N}$.

\subsubsection*{Distributivity check}

Because the operation is defined bilinearly, it satisfies
\[
x*(y+z)=x*y+x*z
\]
and
\[
(x+y)*z=x*z+y*z
\]
for all $x,y,z\in S$.  Hence the operation is both right and left distributive over componentwise addition.

\subsubsection*{Non-associativity check}

Take
\[
a=b=c=e_1.
\]
Then
\[
(e_1*e_1)*e_1=e_2*e_1=e_1.
\]
On the other hand,
\[
e_1*(e_1*e_1)=e_1*e_2=0.
\]
Since
\[
e_1\neq 0,
\]
we have
\[
(e_1*e_1)*e_1\neq e_1*(e_1*e_1).
\]
Thus the operation is non-associative.

Therefore this example lies in the non-associative--distributive quadrant.  It shows that distributivity does not force path-independence.

\subsection{Example 3: Associative but non-distributive toy model}

Let
\[
S=\mathbb{N}
\]
with the usual addition.  Define
\[
a*b=a+b+1.
\]

\subsubsection*{Associativity check}

For all $a,b,c\in S$,
\[
(a*b)*c=(a+b+1)*c=a+b+1+c+1=a+b+c+2.
\]
Similarly,
\[
a*(b*c)=a*(b+c+1)=a+b+c+1+1=a+b+c+2.
\]
Therefore
\[
(a*b)*c=a*(b*c).
\]
Hence the operation is associative.

\subsubsection*{Failure of right distributivity}

Take arbitrary $a,b,c\in S$.  Then
\[
a*(b+c)=a+b+c+1.
\]
But
\[
a*b+a*c=(a+b+1)+(a+c+1)=2a+b+c+2.
\]
These are not equal in general.  For example, with
\[
a=1,\qquad b=0,\qquad c=0,
\]
we get
\[
a*(b+c)=1*(0)=2,
\]
whereas
\[
a*b+a*c=(1*0)+(1*0)=2+2=4.
\]
Thus right distributivity fails.

\subsubsection*{Failure of left distributivity}

Similarly,
\[
(a+b)*c=a+b+c+1,
\]
whereas
\[
a*c+b*c=(a+c+1)+(b+c+1)=a+b+2c+2.
\]
These are not equal in general.  For example, with
\[
a=0,\qquad b=0,\qquad c=1,
\]
we get
\[
(a+b)*c=0*1=2,
\]
whereas
\[
a*c+b*c=(0*1)+(0*1)=2+2=4.
\]
Thus left distributivity also fails.

Therefore this example lies in the associative--non-distributive quadrant.  It shows that path-independence may hold even when mixture additivity fails.

\subsection{Example 4: Non-distributive and non-associative model}

Let
\[
S=\mathbb{N}
\]
with the usual addition.  Define
\[
a*b=a+2b+1.
\]

\subsubsection*{Failure of associativity}

For all $a,b,c\in S$,
\[
(a*b)*c=(a+2b+1)*c=a+2b+1+2c+1=a+2b+2c+2.
\]
On the other hand,
\[
a*(b*c)=a*(b+2c+1)=a+2(b+2c+1)+1=a+2b+4c+3.
\]
These are not equal in general.  For example, take
\[
a=0,\qquad b=0,\qquad c=0.
\]
Then
\[
(a*b)*c=(0*0)*0=1*0=2,
\]
while
\[
a*(b*c)=0*(0*0)=0*1=3.
\]
Hence
\[
(a*b)*c\neq a*(b*c).
\]
Thus the operation is non-associative.

\subsubsection*{Failure of right distributivity}

For all $a,b,c\in S$,
\[
a*(b+c)=a+2(b+c)+1=a+2b+2c+1.
\]
But
\[
a*b+a*c=(a+2b+1)+(a+2c+1)=2a+2b+2c+2.
\]
These are not equal in general.  For example, with
\[
a=0,\qquad b=0,\qquad c=0,
\]
we get
\[
0*(0+0)=1,
\]
whereas
\[
0*0+0*0=1+1=2.
\]
Thus right distributivity fails.

\subsubsection*{Failure of left distributivity}

For all $a,b,c\in S$,
\[
(a+b)*c=a+b+2c+1.
\]
But
\[
a*c+b*c=(a+2c+1)+(b+2c+1)=a+b+4c+2.
\]
These are not equal in general.  For example, with
\[
a=0,\qquad b=0,\qquad c=0,
\]
we get
\[
(0+0)*0=1,
\]
whereas
\[
0*0+0*0=1+1=2.
\]
Thus left distributivity fails.

Therefore this example lies in the non-associative--non-distributive quadrant.  It represents a strongly non-ideal regime in which both path-independence and non-interference fail.

\subsection{Summary of the four examples}

The four examples above may be summarized as follows:
\begin{center}
\scriptsize
\begin{tabular}{@{}c c p{0.25\textwidth} p{0.34\textwidth}@{}}
\textbf{Example} & \textbf{State set} & \textbf{Operation} & \textbf{Regime}\\
\hline
1 & $\mathbb{R}_{\geq 0}$ & $a*b=ab$ & associative and distributive\\
2 & $\mathbb{N}^2$ & bilinear non-associative product & non-associative and distributive\\
3 & $\mathbb{N}$ & $a*b=a+b+1$ & associative and non-distributive\\
4 & $\mathbb{N}$ & $a*b=a+2b+1$ & non-associative and non-distributive
\end{tabular}
\end{center}
These examples confirm that the fourfold regime table is algebraically non-vacuous: each quadrant can occur under explicit symbolic operations.

\bibliographystyle{plainnat}
\bibliography{references}

\end{document}